# Relative Income and Gender Norms:
# Evidence from Latin America

**Ercio Muñoz**
Inter-American Development Bank

**Dario Sansone**
University of Exeter

**João Tampellini**
Vanderbilt University

August 2025

**ABSTRACT**

Using data from over 500,000 dual-earner households in Mexico, we provide evidence of discontinuities in the distribution of relative income within households in Latin America. Similar to high-income countries, we observe a sharp drop at the 50% threshold – where the wife earns more than the husband – but the discontinuity is up to five times larger and has increased over time. These patterns are robust to excluding equal earners, self-employed individuals, or couples in the same occupation/industry. Discontinuities persist across subgroups, including couples with or without children, married or unmarried partners, and those with older wives or female household heads. We also find comparable discontinuities in Brazil and Panama, as well as among some same-sex couples. Moreover, women who are primary earners continue to supply more non-market labor than their male partners, although the gap is narrower than in households where the woman is the secondary earner.

**Keywords:** Gender norms, Relative income, Latin America
**JEL:** D13, D91, J12, J15, J16, O15, Z13

Muñoz: erciom@iadb.org; Sansone: d.sansone@exeter.ac.uk; Tampellini: joao.p.tampellini@vanderbilt.edu. Financial support through the Inter-American Development Bank RG-E2016 is gratefully acknowledged. The views expressed in this paper are those of the authors and should not be attributed to the Inter-American Development Bank. We have benefited from audience participation at Vanderbilt University and University of Exeter. We also thank Josh C. Martin for helpful comments and suggestions. All errors are our own. For the purpose of open access, the authors have applied a Creative Commons Attribution (CC BY NC ND) license to any Author Accepted Manuscript version arising.

# 1. Introduction

A set of papers in economics, following Bertrand, Kamenica, and Pan (2015), identifies a large discontinuity to the right of the 50% mark in the distribution of households according to the wife's share of income in high-income countries. As traditional models of household formation and labor specialization do not predict such gaps, Bertrand and co-authors attribute the discontinuity, at least partially, to gender norms that discourage women from earning more than their husbands. Subsequent papers have expanded the findings to other high-income countries and contested some of the main conclusions (Bertrand et al., 2015; Binder & Lam, 2022; Hederos & Stenberg, 2022; Zinovyeva & Tverdostup, 2021).

Today, however, the majority of the global population lives in low- and middle-income countries, where more conservative gender norms are prevalent and female labor force participation is lower (Inglehart et al. 2014; Verick 2014). In this paper, we use data from Mexico– for which we have more detailed household information– and then expand the analysis to Brazil and Panama to examine the prevalence of these patterns around the world, particularly in areas with more conservative gender norms. Indeed, as shown in Figure A1, a significant proportion of both men and women in these countries hold traditional views on family roles and express negative attitudes toward couples in which the woman earns more than her husband.

Our main analyses are based on the 2015 Mexican Intercensal Survey, including over 500,000 dual-earner households. Our first result confirms the presence of a discontinuity in the distribution of relative income at 0.5. Using the McCrary (2008) discontinuity test, we find that the discontinuity is roughly twice as large as in the U.S. and Finland (Bertrand et al., 2015; Zinovyeva & Tverdostup, 2021) and five times as large as in Sweden (Hederos & Stenberg, 2022). We also document that the discontinuity has grown since 2000. These findings are replicated using census data from Brazil and Panama, highlighting the persistence of such discontinuity over time and across countries.

These results are robust to the exclusion of the mass point of equal earners in the distribution, self-employed individuals, and couples where both individuals work in the same occupation in the same industry. This is in contrast to previous studies in high-income countries arguing that such discontinuities are mainly driven by equal earners and co-working couples (Hederos & Stenberg, 2022; Zinovyeva & Tverdostup, 2021). We then find discontinuities at the 0.5 mark in the distributions of relative income among both couples with and without children, thus suggesting that fertility is not the main determinant of this drop. Remarkably, the discontinuity at 0.5 is found also in less traditional households, such as unmarried couples (potentially reflecting changing trends in cohabitation in Latin America, as documented in Esteve, Lesthaeghe, and López-Gay, 2012), couples in which the woman is the household head, and couples in which the woman is older than her partner.

To better understand the role of intra-household gender differences, we expand the analysis to same-sex couples, where gender norms may be less salient (Andresen & Nix, 2022; Oreffice & Sansone, 2023; Van Der Vleuten et al., 2024). We report the distribution of relative income for younger people versus their older partners, as well as household heads versus their partners. Despite the expected lower influence of these norms, we still find a discontinuity at 0.5 among female same-sex couples. However, the absence of a consistent pattern across countries suggests that the influence of gendered behavioral norms may be context-specific and not universally observed for same-sex households.

We further expand our study by analyzing women's non-market labor supply. We document that the hours gap in non-market labor supplied by the wife and husband is significantly smaller in households where the wife is the primary earner. However, despite the smaller gap, primary earner women are, on average, still the main suppliers of non-market labor for their households, including childcare. Additionally, once the woman becomes the primary earner, increases in the woman's share of household income are associated with a slower convergence between partners in hours of household production. Furthermore, we find that women with higher potential income are more likely to be in the labor force in Mexico, although the effects are small. This contrasts with Bertrand et al. (2015), who argue that women who are likely to outearn their husbands reduce their labor supply to avoid reversing traditional gender norms.

By highlighting the importance of structural factors within households, we contribute to the literature on the importance of gender norms in shaping intra-household dynamics. Previous work has emphasized the role of gender dynamics in shaping marriage market sorting and outcomes (Choo & Siow, 2006; Goñi, 2022; Lundberg & Pollak, 1996; Pollak, 2019; Torche, 2010) and their implications for intra-household bargaining power and labor market allocation (Akerlof &

Kranton, 2000; Bursztyn et al., 2017, 2020; Cavapozzi et al., 2021; Codazzi et al., 2018; Goldin, 2014; Jayachandran & Voena, 2025). However, by showing that relative income discontinuities persist in contexts where traditional gender hierarchies are less clear (e.g., same-sex couples) or disrupted (e.g., female-headed households), our findings suggest that household specialization can emerge from mechanisms beyond norm adherence alone. This highlights the importance of structural factors that may shape intra-household bargaining independently of or in interaction with gender norms.

More broadly, we contribute to the broader literature on household economics (i.e., Greenwood, Guner, and Vandenbroucke 2017; Cortés and Pan 2023), with a particular focus on the intersection of gender norms and the distribution of relative income within couples (Bertrand et al., 2015; Binder & Lam, 2022; Hederos & Stenberg, 2022; Lippmann et al., 2020; Zinovyeva & Tverdostup, 2021), and how this contrasts with traditional Becker's models (Becker, 1973, 1981). While prior research has largely concentrated on high-income countries– particularly in Western Europe and the United States– we extend this analysis to middle-income countries, offering the first systematic, cross-country evidence from Latin America. In doing so, we provide novel insights into how these patterns manifest outside the Global North. Additionally, we bridge this literature with the growing body of work on LGBTQ+ issues in economics (Badgett et al., 2021, 2024; E. Muñoz, Saavedra, et al., 2024; E. Muñoz, Sansone, et al., 2024; E. A. Muñoz & Sansone, 2024; Tampellini, 2024), by being the first to examine the distribution of relative earnings among same-sex couples. Finally, we contribute to the literature on intra-household time allocation (i.e., Aguiar and Hurst 2007; Juhn and McCue 2017) by leveraging one of the richest time-use datasets globally to document how couples' decisions regarding non-market labor are shaped by their relative earnings within the household.

## 2. Data and Sample Construction

### 2.1. Mexican Census Data

For our main analyses, we use the 10% sample of the 2015 Mexican Intercensal Survey (Census data hereafter). The data includes demographic information, including gender, age, and race, as well as various socioeconomic characteristics. Additionally, the 2015 Census was, at the time of its collection, the largest survey in the world to collect information on weekly hours spent on non-market work across several activity categories.

We define different-sex and same-sex couples based on the reported relationship between household members to the head of household. To set up our main samples, we follow Bertrand, Kamenica, and Pan (2015) and restrict the data to households in which both individuals are between 18 and 65 years old and have positive labor income.[1] In Mexico, the household head is defined as "the person recognized as such by the regular residents of the dwelling, through which the bond or kinship relationship of each resident is known to this person. If no one is identified as the head of the dwelling, then the head is considered the first person of reference 12 years old or older who is mentioned by the Informant." We restrict the sample to households in which only one household head and one partner are identified, excluding those that report having more than one partner or more than one household head.[2]

Our measure of income refers to labor earnings and does not include retirement, pension, social programs, transfers, and other sources.[3] While income is technically top-coded, the cap is rarely binding, with only 199 individuals out of over 1 million respondents reporting earnings above the threshold. After making these initial restrictions, our overall sample is composed of 601,682 different-sex couples (433,284 married and 169,398 cohabiting) and 5,915 same-sex couples (3,313 female and 2,602 male). We do not distinguish between married and unmarried same-sex couples, as same-sex marriage had not yet been nationally recognized in Mexico in 2015.

**2.2. The Case of Equal Earners**

For all our main analyses, we make further sample restrictions to make our results comparable to previous literature. Prior work has documented that, in high-income countries, much of the observed discontinuities can be attributed to a mass point at which both partners report identical earnings (i.e., relative income = 0.5; Zinovyeva and Tverdostup 2021; Hederos and Stenberg 2022). In our dataset of married and cohabiting couples, 16% of households fall into this category.

---

[1] The questionnaire of the 2015 Intercensal Survey instructs that income questions should be answered by the individual concerned. However, the census does not record who actually provided the response. Thus, there is no guarantee that reported earnings are self-reported rather than reported by another household member (Muñoz et al., 2024).
[2] We provide a detailed description of our main variables in Online Appendix B.
[3] This differs from previous work in the relative income literature, which includes total household earnings (e.g., Bertrand et al., 2015).

In Appendix Table A1, we show that equal-earning couples are characteristically different from couples in the surrounding relative income ranges, thus supporting previous criticisms: they are twice as likely to earn less than the minimum wage or to be self-employed, and are significantly more likely to work in agriculture, retail, or education.

While the presence of the mass point may have important social and economic underlying causes, there are several potential explanations that are unrelated to within-household gender norms. For example, Binder and Lam (2022) argue that this could reflect an equal-earning norm for a subset of the population and/or frictions in the marriage market that reduce search costs for individuals in certain occupations, while Zinovyeva and Tverdostup (2021) argue that, in the Finnish context, mass points are largely driven by the convergence of earnings among co-working couples. Including these couples could inflate the observed discontinuity for reasons orthogonal to our main research question. Thus, unless stated otherwise, we exclude equal-earning couples from our analyses. Our final sample, excluding equal earners, comprises 511,272 households, of which 503,710 are composed of different-sex couples (362,030 married and 141,680 cohabiting) and 4,851 are composed of married and unmarried same-sex couples (2,726 female same-sex couples and 2,125 male same-sex couples).

Table 1 provides summary statistics for our different samples (excluding equal-earning couples). Cohabiting and same-sex couples tend to be younger and earn less than married couples, with unmarried women cohabiting with a male partner particularly concentrated at the lower end of the income distribution. Same-sex couples are also less likely to have children. Gender gaps in earnings and non-market work are large, with women consistently earning less and contributing significantly more hours to unpaid labor, including childcare. Relatedly, household headship remains highly gendered among different-sex couples, and male partners in different-sex couples are more likely to be older than their female partners.

**Table 1: Summary Statistics**

| Couple type: | Different-sex, married | | Different-sex, cohabiting | | Same-sex, married + cohabiting | |
|---|---|---|---|---|---|---|
| Sex: | Women | Men | Women | Men | Female | Male |
| **Individual-level Variables** | | | | | | |
| Age | 40.005 | 42.405 | 35.520 | 38.211 | 38.718 | 38.261 |
|  | (9.133) | (9.543) | (9.361) | (10.263) | (9.827) | (9.841) |
| High school degree | 0.527 | 0.533 | 0.403 | 0.401 | 0.528 | 0.561 |
|  | (0.499) | (0.498) | (0.490) | (0.490) | (0.499) | (0.496) |
| College degree | 0.314 | 0.301 | 0.179 | 0.171 | 0.297 | 0.335 |
|  | (0.464) | (0.458) | (0.383) | (0.376) | (0.457) | (0.471) |
| Monthly Income | 6,515.85 | 9,404.95 | 5,149.10 | 7,296.91 | 7,674.29 | 8,658.67 |
|  | (8,885.63) | (16,032.37) | (5,549.95) | (12,102.59) | (8,328.85) | (10,766.96) |
| <1 minimum wage | 0.115 | 0.025 | 0.138 | 0.031 | 0.067 | 0.049 |
|  | (0.319) | (0.157) | (0.345) | (0.175) | (0.250) | (0.216) |
| <2 minimum wages | 0.408 | 0.188 | 0.513 | 0.255 | 0.310 | 0.269 |
|  | (0.491) | (0.390) | (0.500) | (0.436) | (0.462) | (0.443) |
| Self-employed | 0.205 | 0.189 | 0.196 | 0.190 | 0.174 | 0.191 |
|  | (0.403) | (0.392) | (0.397) | (0.392) | (0.379) | (0.393) |
| Employer | 0.029 | 0.048 | 0.020 | 0.031 | 0.034 | 0.033 |
|  | (0.169) | (0.214) | (0.139) | (0.173) | (0.181) | (0.179) |
| Household Head | 0.106 | 0.893 | 0.186 | 0.813 | - | - |
|  | (0.308) | (0.308) | (0.389) | (0.389) | | |
| Older Partner | 0.193 | 0.685 | 0.259 | 0.643 | - | - |
|  | (0.395) | (0.464) | (0.438) | (0.479) | | |
| Hours spent on non-market work | 53.978 | 19.017 | 55.255 | 19.119 | 31.994 | 29.074 |
|  | (49.751) | (28.379) | (51.366) | (28.351) | (39.501) | (37.573) |
| Hours spent on childcare | 25.347 | 10.526 | 28.119 | 10.797 | 13.977 | 12.306 |
|  | (39.252) | (21.793) | (41.297) | (22.140) | (28.623) | (27.309) |
| **Household-level variables** | | | | | | |
| Relative income | 0.413 | | 0.415 | | | |
|  | (0.155) | | (0.148) | | | |
| Both self-employed | 0.068 | | 0.066 | | 0.060 | 0.056 |
|  | (0.252) | | (0.248) | | (0.237) | (0.231) |
| Both employers | 0.011 | | 0.006 | | 0.012 | 0.010 |
|  | (0.105) | | (0.078) | | (0.108) | (0.101) |
| Same occupation and industry | 0.075 | | 0.080 | | 0.075 | 0.064 |
|  | (0.264) | | (0.272) | | (0.263) | (0.245) |
| Age gap | 3.586 | | 5.127 | | 4.399 | 4.370 |
|  | (3.579) | | (4.976) | | (4.493) | (4.630) |
| Any children in HH | 0.743 | | 0.714 | | 0.632 | 0.525 |
|  | (0.436) | | (0.451) | | (0.482) | (0.499) |
| Any children <5 in HH | 0.272 | | 0.344 | | 0.239 | 0.214 |
|  | (0.445) | | (0.344) | | (0.482) | (0.410) |
| Households | 362,030 | | 141,680 | | 2,726 | 2,125 |

*Notes*: Weighted statistics. See Online Appendix B for detailed variable description. Source: 2015 Mexican Intercensal Survey. "<1 minimum wage" and "<2 minimum wage" are indicators equal to one if the individual earns, respectively, less than the monthly minimum wage or less than two times the monthly minimum wage. For same-sex couples, we omit variables that require ordering individuals within the couple (e.g., identifying the older partner or household head), as there is no clear way to assign such roles. Standard deviation in parentheses.

## 3. Documenting Patterns in Mexico

### 3.1. Graphical Evidence of Discontinuities

Following Bertrand, Kamenica, and Pan (2015), we define, for a different-sex household $h$, $relIncome_h = \frac{wifeIncome_{ih}}{wifeIncome_{ih} + husbIncome_{ih}}$, where $i$ indexes individual income. We first present results for married different-sex households. Figure 1 shows the distribution of households in Mexico according to the wife's share of household income. In our sample, 26% of wives earn more than their husbands. We divide households into 20 bins centered at the midpoint of the bin and present the distribution along with a LOWESS (locally weighted scatterplot smoothing). Panel A shows the distribution including equal-earning couples, and Panel B shows the distribution of couples for the sample used throughout our analysis.

These figures are close to Figure I in Bertrand, Kamenica, and Pan (2015) for the US. As documented in previous studies (i.e., Zinovyeva and Tverdostup 2021; Hederos and Stenberg 2022), the modal share of the household labor income earned by the wife in Mexico is between 35-40%. Similarly, as reported in Table 1, the average wife's relative income in this group is 41.3%.

As in Bertrand, Kamenica, and Pan (2015), the main finding is the sharp drop to the right of 0.5 in the distribution of the share of total household labor earnings earned by the wife, even without considering equal-earning households. In other words, it is much less likely for a wife to earn just above her husband's earnings than just below her husband's earnings.

To address potential concerns about rounding or bunching around salient thresholds, Figure A2 presents a kernel density of the wife's share of household income. While the use of survey data introduces non-smoothness in reported earnings unlike administrative tax data used in other studies (i.e., Zinovyeva and Tverdostup 2021; Hederos and Stenberg 2022), we still observe a pronounced discontinuity at 0.5. In Figure A3, we replicate Figure 1 after adding uniformly distributed random noise of ±1% to each individual's income. The noise was only added to households where relative income is not exactly 0.5, as equal-earning couples may be characteristically different. This noise smooths out small rounding or heaping artifacts in the data, helping ensure that observed discontinuities– such as the drop at 0.5 in the wife's share of household income– are not mechanical results of data irregularities. The persistence of the discontinuity under this approach suggests that, despite differences in data structure from previous studies, similar behavioral mechanisms may be at work.

**Figure 1: Distribution of the share of total household labor earnings by the wife in Mexico**

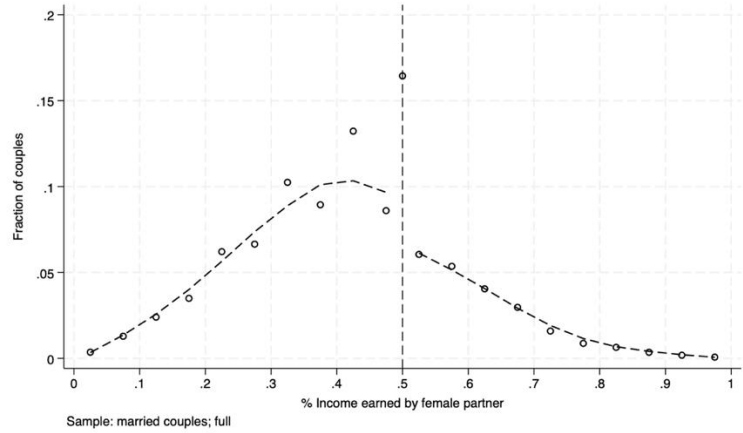

(a) Including Equal Earners

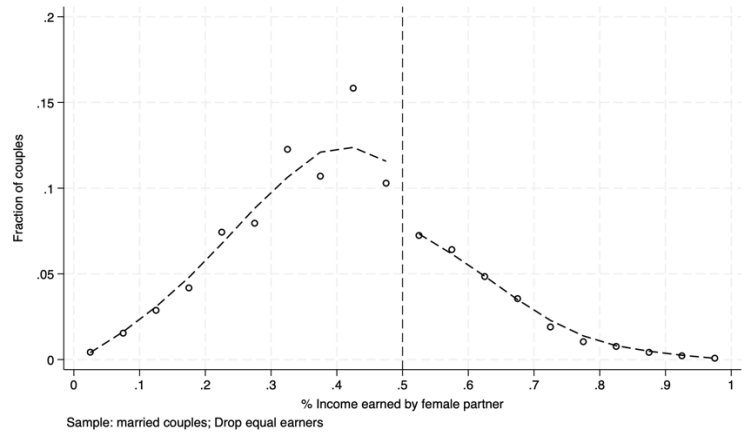

(b) Excluding Equal Earners

*Notes:* Data from the 2015 Mexican Intercensal Survey. The sample is restricted to different-sex married couples in which both spouses are between 18 and 65 years old and report positive labor income. Couples with identical reported incomes are excluded in Panel (b). Each point represents the share of couples within a 0.05-wide bin of relative income. The vertical line marks the 0.5 threshold, and the curve shows a LOWESS-smoothed fit to the distribution, allowing for a potential discontinuity at 0.5.

### 3.2. Testing for Size and Statistical Significance

We use the McCrary (2008) test for discontinuities in log difference in heights. Let be $s_{ijc}$ be the share of the household income earned by the female partner $i$ in couple $j$, with $s_{ij} \in (0,1]$. We analyze the distribution of this variable using its cumulative distribution function, $F(s) = \Pr(S \leq$ $s$), and the associated probability density function $f(s) = \frac{dF(s)}{d(s)}$. We use the automatically selected bin and bandwidth provided by the test (McCrary, 2008). For the cutoff point $c = 0.50001$, the coefficient of interest $\theta$ is given by:

$$\theta = \log \lim_{s \downarrow c} f(s) - \log \lim_{s \uparrow c} f(s) \equiv \log f^+ - \log f^- \tag{1}$$

We first quantify the discontinuity shown in Figure 1, restricting our sample to married couples. Table 2 shows the resulting estimates. As shown in Column 1, we estimate a sharp drop of 21.6 log points at the 0.50001 threshold in the distribution of households according to relative income, statistically significant at the 1 percent level. This implies that, among households near the cutoff, those where the wife earns just more than half the total income are about 19.4% less common than those just below the threshold – reflecting a sharp drop in the density of the relative income distribution at 0.5. This estimate is robust to using alternative slightly different cutoffs ($c = 0.5$ and $c = 0.49999$), as shown in Table A2.

These discontinuities are robust to dropping couples where both individuals are self-employed and where both individuals report working in the same occupation and industry (Table 2 Columns 2-4). This extension is noteworthy, as these groups were the main drivers of the discontinuities previously documented in the literature for Europe and the US (i.e. Hederos & Stenberg, 2022; Zinovyeva & Tverdostup, 2021). The coefficient for the discontinuity becomes even larger when employers and self-employed individuals are excluded from the sample, and smaller (but still larger and significant) when couples working in the same industry and occupation are excluded. This latter group is particularly relevant, as spouses working in the same occupation and industry are considerably more likely to share a workplace (Hyatt, 2019), which may introduce workplace-based constraints on household decision-making. In previous studies, such shared work environments were shown to reinforce social comparisons or peer effects that amplify norm-driven behaviors. The fact that we continue to find strong discontinuities even after removing these cases suggests that the patterns we observe are not merely artifacts of shared professional environments but reflect deeper household-level dynamics.

Overall, we find that the discontinuities documented in Mexico are 4-5 times larger than previously documented in Sweden and 2 times larger than previously documented in Finland and the United States. It is worth emphasizing that these estimates are obtained after excluding equal-earning couples from the sample, so unlike previous findings in North America and Europe (Bertrand et al., 2015; Hederos & Stenberg, 2022; Zinovyeva & Tverdostup, 2021), the results are not driven by the presence of a mass point at the midpoint of the distribution.

**Table 2: McCrary Test for discontinuity in the distribution of the woman's share of total household labor income – Different-Sex Married Partners**

| | All samples below exclude equal-earning couples | | | |
|---|---|---|---|---|
| | Full Sample | Excluding employers/self employed | Excluding couples in same occupation and industry | Excluding employers/self emp./same occ. and industry |
| | (1) | (2) | (3) | (4) |
| **Panel A: 2015** | | | | |
| log distance at 0.50001 | -0.216*** | -0.245*** | -0.175*** | -0.200*** |
| | (0.010) | (0.012) | (0.010) | (0.013) |
| $N$ | 362,030 | 211,517 | 331,978 | 191,820 |
| **Panel B: 2000** | | | | |
| log distance at 0.50001 | -0.176*** | -0.207*** | -0.142*** | -0.170*** |
| | (0.013) | (0.014) | (0.013) | (0.014) |
| $N$ | 231,118 | 164,691 | 214,303 | 153,027 |

*Notes:* *** $p<0.01$ ** $p<0.05$ * $p<0.10$. Each cell reports the estimated discontinuity in the density of the woman's share of total household labor income at the 50% threshold, using the McCrary (2008) test. The dependent variable is the log difference in the density just above versus just below the 50% threshold. All samples exclude couples with exactly equal earnings. Panel A uses data from the 2015 Mexican Intercensal Survey, and Panel B uses the 2000 Mexican Census. Column (1) includes the full sample of different-sex married couples. Column (2) excludes self-employed and employer couples. Column (3) excludes couples where both spouses work in the same occupation and industry. Column (4) excludes both self-employed/employers and couples in the same occupation and industry. Standard errors are in parentheses.

Further, these discontinuities have grown over time. Using data from the 2000 Mexican Census and applying the same sample restrictions as our main analysis, we show in Panel (a) of Figure A4 and Panel B of Table 2 that the gap has actually grown from 17.6 to 21.6 percent in 15 years. Results for the 2010 Census, in Panel (b), show a magnitude similar to 2000. In Panel (c) of Figure A4, we include the figure for the 2020 Mexican Census, in which the magnitude of the discontinuity remains similar. We note, however, that at the time of writing, the publicly available version of the 2020 data does not include fully harmonized occupation and income information.

## 3.3. Heterogeneity by Family Composition

We then explore whether differences in the structure of dual-earning families help explain the observed discontinuities in the distribution of households by the wife's share of labor income. Specifically, we look at heterogeneity by marital status and the presence of children.

Children may influence incentives around relative income in various ways, especially given the large literature on the motherhood penalty and fatherhood premium (Goldin, 2014; Kleven et al., 2019; Waldfogel, 1998): according to the Child Penalty Atlas, Mexico has one of the largest child penalty for women in the world (Kleven et al., 2024). Nevertheless, it is important to note that Mexico does not offer tax benefits for married couples: individuals are not allowed to file jointly regardless of marital or parental status. In Panels (a) and (b) of Figure 2, we show that the distributions of relative income in households with and without children are similar. In Table A3, we show that the coefficients for the McCrary test are larger in magnitude for households with children, although sizeable discontinuities are also observed for households without children. Overall, it is unlikely that parental status is the main driver of these gaps, even though women are disproportionately responsible for childcare, even when they are the breadwinners (see Section 5.3 for a more detailed discussion on time use).

We then turn to unmarried, cohabiting couples. Using data from the Netherlands, Kalmijn, Loeve, and Manting (2007) showed a heterogeneous impact of relative income on household dynamics: for married couples, increasing the wife's share of household earnings was associated with an increase in the probability of divorce for married couples, while the opposite was true for cohabiting couples. While we are unable to observe marital dynamics over time, we show in Panels (c) and (d) of Figure 2, as well as Table A4, that the discontinuities for cohabiting couples are larger than those observed for married couples.

While initially counterintuitive, as cohabiting couples are already breaking traditional social expectations by living together before marriage, such dynamics may reflect recent cohabitation patterns in Latin America, where cohabitation has become increasingly common among individuals with lower income and educational levels (Esteve et al., 2012). Indeed, Table A5 shows that, since at least 2000, cohabiting couples in Mexico have become increasingly negatively selected relative to those who marry in terms of income and educational attainment.

Therefore, low-educated low-income cohabiting couples may actually hold stronger gender norms than married couples, which would explain the observed patterns.

**Figure 2: Distribution of relative earnings according to family structure**

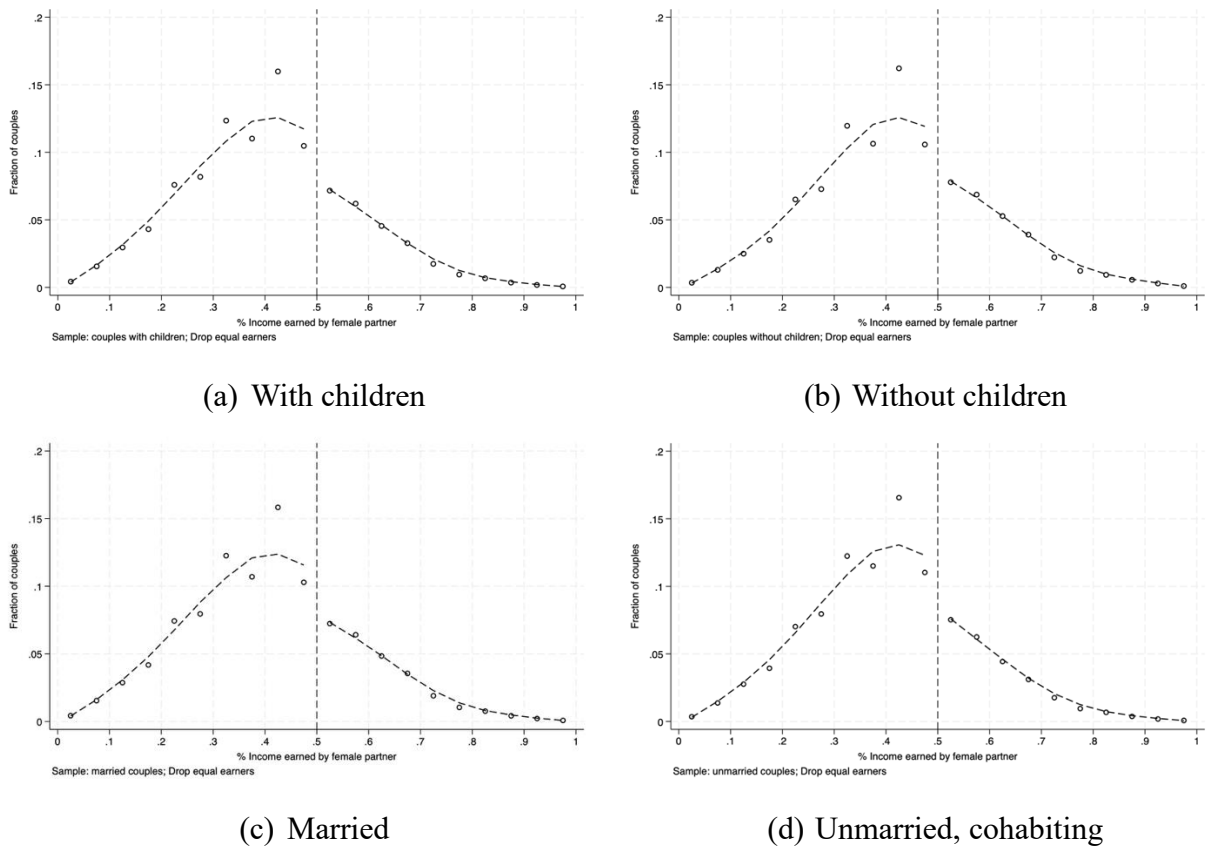

(a) With children  (b) Without children

(c) Married  (d) Unmarried, cohabiting

*Notes:* Data from the 2015 Mexican Intercensal Survey. The sample is restricted to different-sex couples in which both individuals are between 18 and 65 years old and report positive labor income. Further sample restrictions are described in each subfigure caption. Couples with identical reported incomes are excluded. Each point represents the share of couples within a 0.05-wide bin of relative income. The vertical line marks the 0.5 threshold, and the curve shows a LOWESS-smoothed fit to the distribution, allowing for a potential discontinuity at 0.5.

## 4. Sorting, Preferences, and Other Household Dynamics

In this section, we explore a range of intra-household dynamics that may contribute to discontinuities in the distribution of relative income. Indeed, Binder and Lam (2022) argue that such discontinuities are shaped not only by gender norms but also by marriage market sorting – and by the underlying distributions of traits correlated with income that couples sort on. Thus, even in the absence of explicit gender norms, gaps can emerge if people consistently match on characteristics related to income. While we are unable to observe couple formation *ex-ante*, these decisions may reflect underlying power dynamics or economic considerations that, once a couple is formed, interact with household labor decisions in ways that shape relative income.

To provide an intuitive exercise about the potential role of sorting, we simulate a counterfactual in which individuals are randomly matched to different-sex partners, thus removing any systematic sorting by income or other traits. In Figure A5, we present the average distribution of relative income across 100 random reassignments, showing that the sharp discontinuity fully disappears. While this exercise helps us rule out that our results are artifacts of the data structure, we emphasize that random matching does not account for underlying differences in the distribution of traits along which couples often sort (e.g., education, income; Binder and Lam 2022). As such, we interpret it primarily as a plausibility check than evidence on matching behavior. These results can also be viewed as a placebo test. Specifically, once individuals are randomly matched, there is no discontinuity at 0.5. This suggests that the discontinuity observed in Figure 1 is not something that would arise by chance, but rather reflects deeper mechanisms beyond random noise in the data.

### 4.1. Household Head Dynamics

One intra-household dimension that may influence how couples make labor decisions is who is designated as the household head. In our data, the head is identified at the time of enumeration, and this designation may reflect social expectations (or administrative convenience). As shown in Table 1, men are overwhelmingly identified as the household head in different-sex couples, consistent with traditional views of male authority within the household. We provide additional detail on how this variable was collected in Appendix B.

If household headship is correlated with bargaining power or normative expectations, it could influence labor supply decisions and, in turn, the distribution of relative income. To assess this, we test for discontinuities separately for couples where the wife is identified as the household head. These households may already depart from conventional gender roles, so differences in the distribution of relative income may clarify whether the observed discontinuities reflect broader social dynamics.

Figure 4 shows the distribution of households according to the wife's share of income in households where she is identified as the household head. While these couples may appear to challenge conventional gender roles, we still observe a clear discontinuity at the 50% threshold. This suggests that the designation of household headship does not necessarily translate into higher relative earnings or more egalitarian household dynamics. Previous work has shown that even when women hold formal markers of authority, deeply rooted social norms can still shape labor and income decisions (Gu et al., 2024; Ke, 2021). Thus, in our context, household headship may reflect administrative formality more than actual changes in bargaining power.

Relatedly, Panel (a) in Figure A6 presents the distribution of households ordered by the share of income earned by the non-household head, regardless of gender. We find a clear discontinuity at the 50% threshold, similar in magnitude to our main results. Given that 89% of households in our sample are headed by men, this alternative ordering largely mirrors the dynamics of female earners in male-headed households.

**Figure 3: Distribution of households according to relative income in households where the wife is the household head**

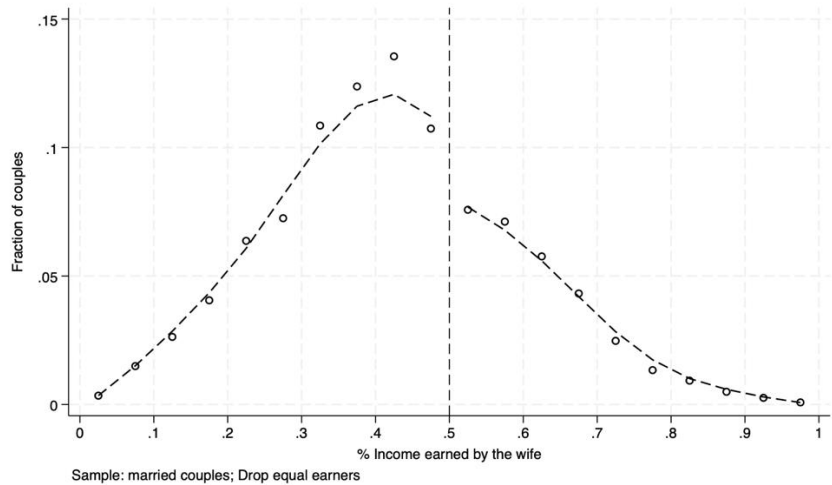

*Notes:* Data from the 2015 Mexican Intercensal Survey. The sample is restricted to different-sex married couples in which both spouses are between 18 and 65 years old, report positive labor income, and the wife is identified as the household head. Couples with identical reported incomes are excluded. Each point represents the share of couples within a 0.05-wide bin of relative income. The vertical line marks the 0.5 threshold, and the curve displays a LOWESS-smoothed fit to the distribution.

### 4.2. Differences in age

Another intra-household factor that may shape earnings dynamics is the age differences between partners. In our different-sex couple samples, men are older than their partners, a pattern often linked to traditional social norms. When this pattern is reversed, the couple may already deviate from conventional gender roles, potentially affecting the salience of gender norms.

To explore this, we test for discontinuities in the distribution of relative income for couples where the woman is the older partner by 5 years or more. If the observed drop at the 50% mark is largely driven by traditional gender roles, we would expect the discontinuity to be smaller for these households. Figure 5 shows that this is not the case, as the discontinuity persists, although smaller in magnitude. This suggests that while age dynamics may interact with gendered expectations, they are not sufficient to fully eliminate broader structural patterns shaping relative income within households.

**Figure 4: Distribution of households according to relative income in households where the wife is 5+ years older than her husband**

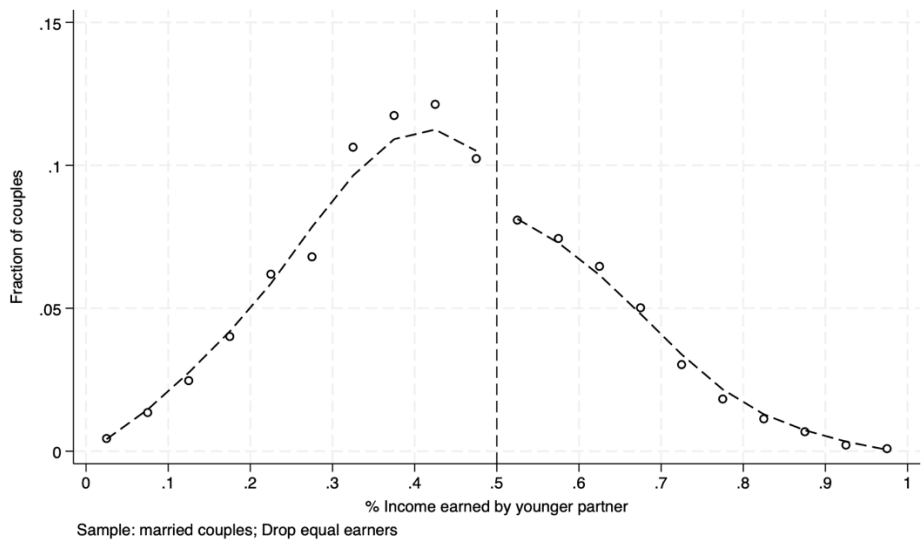

*Notes:* Data from the 2015 Mexican Intercensal Survey. The sample is restricted to different-sex married couples in which both spouses are between 18 and 65 years old, report positive labor income, and the wife is at least five years older than her husband. Couples with identical reported incomes are excluded. Each point represents the share of couples within a 0.05-wide bin of relative income, measured as the share earned by the younger partner. The vertical line marks the 0.5 threshold, and the curve displays a LOWESS-smoothed fit to the distribution.

Nevertheless, we show in Figure A7 that the observed discontinuities become smaller as the age gap between partners increases. For couples where the wife is at least ten years older than her husband, the discontinuity disappears entirely. These couples likely differ in both social dynamics and incentives surrounding labor decisions, as partners are likely in different stages of their careers. One interpretation is that reversing the typical age hierarchy helps in offsetting the effects of assortative matching, where couples tend to form based on traits that reinforce traditional gender expectations. In these less conventional couples, traditional expectations around income and household roles may be less salient, leading to smoother relative income distributions.

Relatedly, in Panel (b) of Figure A6, we show a sizable discontinuity when households are ranked by the income share of the younger partner, regardless of gender. Given that men are the older partner in 68% of households, this pattern is consistent with the idea that relative age– like household headship– may proxy for bargaining power or normative expectations.

### 5. The Case of Same-Sex Couples

So far, our results point to large discontinuities in the distribution of households by the female partner's earnings. These discontinuities have been persistent over time and cannot be fully explained by differences in marital status, marriage type, or the presence of children. To better understand the role of intra-household gender norms in shaping these patterns, we next turn to same-sex couples – a comparison that has been largely underexplored in the relative income literature. Previous studies have used same-sex couples to understand the role of biology and gender dynamics in household decision-making, suggesting that intra-household gender bargaining may be less relevant for same-sex households (i.e., Andresen and Nix 2022; Oreffice and Sansone 2023; Van Der Vleuten, Evertsson, and Moberg 2024; Moberg and Van Der Vleuten 2025).

As described in the data section, we identify same-sex couples using the *relationship to household head* variable, following previous work (Badgett et al., 2021; E. Muñoz, Sansone, et al., 2024). Specifically, we define a couple as same-sex if the person listed as the household head's partner (married or unmarried) is of the same sex as the household head.[4] We then apply the same sample restrictions as described in Section 2.

---

[4] We make no distinction between married and unmarried cohabiting partners in our sample of same-sex households since same-sex unions were not nationally recognized in Mexico as of 2015. Unlike the United States, which forcibly recoded same-sex couples as different-sex couples until 2008 (Badgett et al., 2021), we still observe 3,537 same-sex couples who report being married (2,001 female and 1,536 male).

By construction, couples in our sample are composed of individuals of the same sex. As such, we are unable to present discontinuities along the distribution of households according to the *wife's* share of household earnings. Therefore, our main results in this section present discontinuities along the share of income earned by the *non-household head* member. In Table A6, we also present discontinuities for the distribution of income earned by the *younger* partner. While these comparisons are not directly equivalent to the ones used for different-sex couples, we note that the husband is the household head in 89% and the older partner in 69% of households in our sample of different-sex households. Still, as discussed in the previous sections, we show in Figure A6 that substantial discontinuities remain in the different-sex married sample when we use the same measures described in this section: the younger partner is less likely to earn just above their partner's earnings than just below it. The same is true for the non-household heads in different-sex couples. Again, this is in line with the main results in Figure 1, as the vast majority of the younger partners and non-household heads in different-sex couples are women.

We show our main results for same-sex couples in Figure 6, where we detect sizeable discontinuities for female same-sex couples in Mexico, but not for male same-sex couples. In Figure A8, we show that the results are qualitatively unchanged when households are ordered by the share of income earned by the younger partner instead of focusing on non-household heads. Table A6 presents the corresponding McCrary test results, where we show a 25 log points gap for female same-sex households in Mexico. We note that, in this context, limited sample sizes – fewer than 3,000 households per subgroup – can limit the interpretation of estimates obtained for density-based statistical tests. A large literature has documented that local polynomial density estimators may underperform in small samples, particularly when there is bunching at specific values or limited support near the cutoff (Cattaneo et al., 2020; Kuehnle et al., 2021; Rosenberg, 2024). As such, we interpret these estimates with caution and place greater emphasis on the visual patterns.

The inexistence of a discontinuity among male same-sex couples in Mexico aligns with the hypothesis of gender norms being less salient for sexual minorities. However, the presence of a discontinuity for female same-sex couples raises the possibility that factors beyond gender norms may shape household income dynamics. One possibility is that gender norms are not the primary driver of these patterns. Previous work shows that same-sex couples tend to specialize less than different-sex couples, in line with Beckerian expectations, though the gap has narrowed over time (Giddings et al., 2014; Hofmarcher & Plug, 2022). The observed discontinuities may, instead, reflect structural differences in household bargaining or other unobserved preferences that shape labor decisions, even outside of traditional gender roles.

**Figure 5: Distribution of the share of total household labor earnings by the non-household head in same-sex couples**

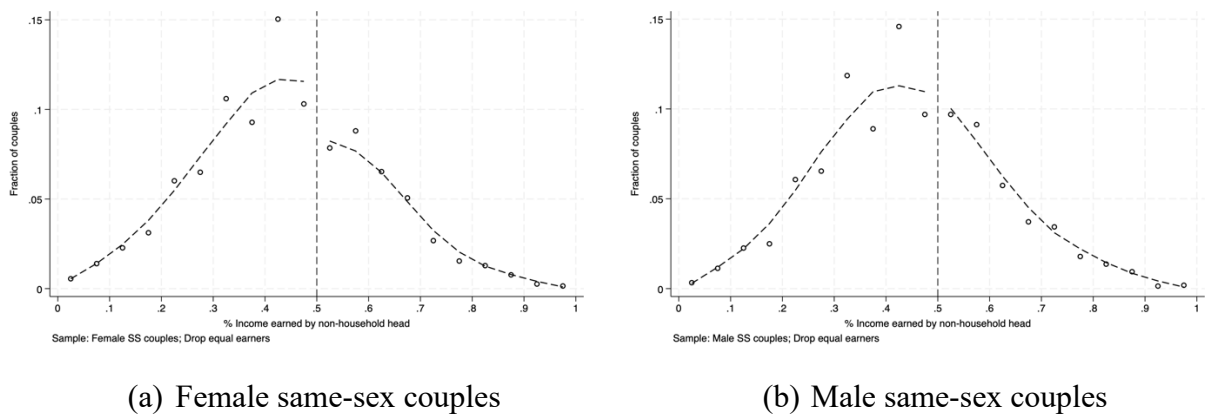

(a) Female same-sex couples          (b) Male same-sex couples

*Notes:* Data from the 2015 Mexican Intercensal Survey. The sample includes same-sex couples where both partners are between 18 and 65 years old and report positive labor income. Couples with equal reported incomes are excluded. Relative income is measured as the share of total household labor earnings contributed by the non-household head. Each point represents the share of couples within a 0.05-wide bin of relative income. The vertical line denotes the 0.5 threshold, and the curve shows a LOWESS-smoothed fit to the distribution. Panels (a) and (b) present results separately for female and male same-sex couples.

Alternatively, the inconsistent pattern of discontinuities may reflect varying preferences for more egalitarian household labor allocations. Prior work has suggested that sexual minorities may hold stronger egalitarian preferences than their heterosexual counterparts (Ciscato et al., 2020). For instance, Verbakel & Kalmijn, (2014) argue that same-sex couples tend to sort more strongly on educational attainment, and that the similarity in schooling levels can facilitate a more equal division of labor within the household. Couples who prioritize shared participation in the labor market may be thus less likely to develop large differences in earnings.

Another possibility is that some same-sex couples may, to some extent, reproduce gendered norms even absent gender differences. Previous work has shown that roles within same-sex households can resemble traditional gender arrangements, with partners sometimes differentiating along lines of gender expression (Badgett, 1995; Doan & Quadlin, 2019; Lamos, 1995; Oreffice, 2011; Van Der Vleuten et al., 2024). This could lead to patterns of specialization that resemble those in different-sex households. While speculative, this interpretation is consistent with the idea that social norms may be internalized even in the absence of gender differences.

Importantly, we note that these interpretations are not mutually exclusive. While we are unable to fully disentangle these mechanisms with the available data, the findings reinforce the central theme of this paper: that gender norms and their influence on household decisions are complex, and shaped by both observable and unobservable traits – even where gender differences may be less salient.

## 6. Relative Income and Labor Supply

So far, we have shown that discontinuities in the distribution of households by the wife's share of income are widespread, significantly larger than previously documented, and can be partially explained by observable household characteristics or matching patterns. In this section, we turn to labor supply decisions within the household to better understand how relative earnings relate to how couples divide their time– especially with respect to labor force participation.

### 6.1. Labor Force Participation

We conduct an analogous exercise to Section IV in Bertrand, Kamenica, and Pan (2015). Here, we are interested in understanding whether couples in which the wife's potential labor market earnings exceed those of the husband respond by simply removing the wife from the labor market, which would represent a strong form of conformity to traditional gender norms.

For this exercise, we expand the sample to include couples where the wife has zero labor income, while still requiring the husband to have positive labor income. We estimate each woman's potential labor market earnings following Bertrand, Kamenica, and Pan (2015). We begin by assigning women to demographic groups based on ethnicity (non-indigenous and indigenous), five-year age bins, education level, and state of residence. Then, for each woman $i$, in household $h$, belonging to group $g$, we compute the $p$th percentile of earnings for women with positive labor income in group $g$, denoted as $w_{i,g}^p$, where $p \in \{5, \dots, 95\}$. Then, for a husband with observed income $husbIncome_h$, we define $PrWifeEarnsMore_{ih} = \frac{1}{19}\sum_p \mathbf{1}\{w_{i,g}^p > husbIncome_h\}$, which reflects the likelihood that a woman with her demographic profile would outearn her husband based on the earnings distribution of similar working women. The mean of $PrWifeEarnsMore_{ih}$ is 0.289 for the sample, similar to the value reported by Bertrand, Kamenica, and Pan (2015) for the 2010 ACS in the United States. Then, we estimate the following model:

$$wifeLFP_{ih} = \alpha_0 + \beta_1 PrWifeEarnsMore_{ih} + \beta_2 \log husbIncome_h \\ + w_{i,g}^p \Phi + X_{ih}\Gamma + \varepsilon_{ih} \qquad (2)$$

Where $w_{i,g}^p \Phi$ are controls for the wife's potential income at each of the vigintiles, and $X_{ih}\Gamma$ include controls for the wife's and husband's education, race, and state of residence fixed effects. Standard errors are clustered at the woman's demographic group level. The coefficient, $\beta_1$, as displayed in the table, represents the effect of moving from 0 to 1 probability that the wife would outearn her husband. When discussing the results below, we scale the coefficient to a one-standard deviation change for ease of interpretation.

Table 3 presents the obtained estimates. In the baseline specification, we find that a one standard deviation increase in the probability that the wife would outearn her husband is associated with a 1.1 percentage point decrease in her probability of participating in the labor force. The negative coefficient, while consistent with the idea that couples may conform to traditional gender norms, is 75% smaller in magnitude than the effect estimated by Bertrand, Kamenica, and Pan (2015).

In column 2, we introduce a cubic polynomial in the log of the husband's earnings to flexibly control for household income. In doing so, the obtained estimate becomes positive. One interpretation is that, absent these controls, the negative association between the wife's earning potential and labor force participation partially reflects that women married to lower-earning husbands are both more likely to outearn them and more likely to stay out of the labor force for financial or other reasons. Once we account for the non-linear relationship between husband's income and wife's labor supply, the residual variation in the wife's earning potential is positively associated with her labor market participation, potentially reflecting stronger labor market attachment among higher potential earnings. Figure A9 supports this interpretation, showing a strong positive relationship between labor force participation rates and the median income within a demographic group. Still, while statistically significant, this relationship is relatively small: a standard deviation increase in the likelihood of a woman outearning her husband is associated with an increase of 0.3 percentage points (1.3% of the mean) in her likelihood of participating in the labor force. In columns 3 and 4, we further include controls for the presence of children in the household, and the coefficients remain relatively stable, further suggesting that parenting status is not a main determinant of labor decisions in this context.

**Table 3: Potential Income and Female Labor Force Participation**

|  | (1) | (2) | (3) | (4) |
| --- | --- | --- | --- | --- |
| *PrWifeEarnsMore* | -0.041*** | 0.013*** | 0.016*** | 0.009** |
|  | (0.003) | (0.004) | (0.004) | (0.004) |
| Mean of *wifeLFP* | 0.283 | 0.283 | 0.283 | 0.283 |
| S.D. of *PrWifeEarnsMore* | 0.285 | 0.285 | 0.285 | 0.285 |
| *N* | 2,351,606 | 2,351,606 | 2,351,606 | 2,351,606 |
| $R^2$ | 0.119 | 0.120 | 0.121 | 0.129 |
| *Additional controls:* |  |  |  |  |
| Cubic in *logHusbIncome* |  | X | X | X |
| Children in household |  |  | X | X |
| Children under 5 in household |  |  |  | X |

*Notes:* *** $p<0.01$ ** $p<0.05$ * $p<0.10$. This table reports the relationship between the wife's predicted likelihood of outearning her husband and her labor force participation. The sample includes married couples in which the husband has positive labor income. The variable *PrWifeEarnsMore* reflects the fraction of percentile earnings thresholds at which a woman with a given demographic profile would outearn her husband, based on the distribution of labor earnings among similar working women. The dependent variable is an indicator for whether the wife participates in the labor force. All specifications control for wife's and husband's education, race, and state fixed effects. Columns 2–4 sequentially add controls: a cubic polynomial in log husband income; indicators for whether the couple has children in the household; and whether they have children under 5. Standard errors are clustered at the woman's demographic group level.

Importantly, we note that the average labor force participation in our sample among married women, 28%, is significantly smaller than the 78% reported in Bertrand, Kamenica, and Pan (2015), which could affect the interpretation of results.[5] Notably, the earnings distribution used to calculate potential income is estimated using realized income among working women, which may not capture the unobserved characteristics of non-working women in the same group. In contexts of low female labor force participation, such as Mexico, selection into the labor market likely reflects differences in preferences and, potentially, gender norms (i.e., Cavapozzi, Francesconi, and Nicoletti 2021). This would likely lead to an upward-biased measure of potential income for non-working women, and, thus, increase the likelihood that non-working women would outearn their husbands. This may explain why we only observe minor effects of predicted relative income on labor force participation.

---
[5] The labor force participation rate for married women in our sample is similar to that estimated by Bhalotra and Fernández (2024).

**6.2. Gap between potential and realized income**

If couples make labor and earnings decisions influenced by gender norms, another possible margin of adjustment may be to reduce the wife's earnings so that she earns less than her husband. To understand this, we define for a woman $i$ in household $h$ a variable $incomeGap_{ih} = \frac{wifeIncome_{ih} - wifePotential_{ih}}{wifePotential_{ih}}$, where $wifePotential_{ih}$ is the mean of the potential income distribution for a woman's demographic group, as described in the previous section. This variable captures distortions in actual earnings relative to expected earnings among women who are active in the labor market. To minimize the influence of extreme outliers, we restrict the sample to women between the 1$^{st}$ and 99$^{th}$ percentiles of the $incomeGap_{ih}$ distribution. The analysis in this section follows the same structure as Section 5.1. The outcome is $incomeGap_{ih}$ instead of $wifeLFP_{ih}$.

The estimate obtained from the baseline specification, shown in column 1, shows that a standard deviation increase in the probability that a wife would outearn her husband increases the gap between her realized and potential incomes by 4.8 percentage points. This estimate is similar in magnitude to that presented by Bertrand, Kamenica, and Pan (2015) using data from the 2008-2010 ACS. However, once we add controls for a cubic polynomial in the husband's income and the presence of children in the household, the coefficient becomes smaller: a one-standard deviation increase in the probability of a wife outearning her husband increases the gap by 3.3 percentage points.

This measure of income gap, which follows directly from Bertrand, Kamenica, and Pan (2015), is based on the mean of the potential income distribution within each demographic group. As income distributions are typically right-skewed, using the mean may overstate the potential earnings for many women, particularly in lower-income groups. Thus, the income gap would be mechanically inflated, especially among those with below-median earnings. Thus, in Table A7, we present an alternative version where the income gap is measured according to the distance to the median income within a woman's demographic group. We note that, on average, the gap between a woman's earning and the median income for her demographic group is positive. Thus, the coefficient on *PrWifeEarnsMore* becomes smaller in magnitude and becomes statistically insignificant when we add controls for the presence of children in the household.

**Table 4: Potential Income and Wife's Realized Earnings**

|  | (1) | (2) | (3) | (4) |
| --- | --- | --- | --- | --- |
| *PrWifeEarnsMore* | -0.167*** | -0.123*** | -0.114*** | -0.115*** |
|  | (0.008) | (0.008) | (0.008) | (0.008) |
|  |  |  |  |  |
| Mean of *incomeGap* | -0.045 | -0.045 | -0.045 | -0.045 |
| S.D. of *PrWifeEarnsMore* | 0.285 | 0.285 | 0.285 | 0.285 |
| N | 613,990 | 613,990 | 613,990 | 613,990 |
| $R^2$ | 0.144 | 0.149 | 0.152 | 0.153 |
|  |  |  |  |  |
| *Additional controls:* |  |  |  |  |
| Cubic in *logHusbIncome* |  | X | X | X |
| Children in household |  |  | X | X |
| Children under 5 in household |  |  |  | X |

*Notes:* *** $p<0.01$ ** $p<0.05$ * $p<0.10$. This table reports the relationship between the wife's predicted likelihood of outearning her husband and the gap between her realized and potential earnings. The dependent variable is defined as the difference between observed and predicted earnings (based on demographic group averages), normalized by potential earnings. The sample is restricted to women in the labor force and excludes extreme outliers by keeping observations between the 1st and 99th percentiles of the outcome variable. All specifications include controls for wife's and husband's education, race, and state fixed effects. Columns 2–4 sequentially add a cubic polynomial in log husband income, indicators for whether there are children in the household, and whether any are under age 5. Standard errors are clustered at the woman's demographic group level.

### 6.3. Non-market Labor Supply

In this section, we explore the relationship between relative income within households and the provision of non-market labor. The 2015 Mexican Intercensal Survey asked, for the first time, the distribution of hours spent on non-market activities across several categories. We provide a detailed description of each category in Appendix B. We focus on two outcomes: total non-market work (including childcare) and hours spent on childcare for households with children. We use the same sample as in our main analysis, which includes only different-sex married couples where both partners have positive labor earnings and excludes equal-earning couples. In Figure A10, we show that our analysis remains qualitatively unchanged by including equal-earning couples, although it is worth noting that both men and women in equal-earning couples spend less time in non-market work than those around the 0.5 threshold (as expected if they are both working full-time).

We begin by documenting patterns in the distribution of hours supplied in non-market labor according to the wife's relative earnings. In Panels (a) and (b) of Figure 7, we show that there are no discontinuities at the 0.5 threshold in the number of hours of non-market labor provided by the female partner. For most of the relative income range, the wife's non-market labor supply is monotonically decreasing, reversing the pattern only when she earns more than 90% of the household's income. For husbands, we show a clear change in slope at the 0.5 threshold. While the husband is the primary earner, his non-market labor supply monotonically increases as the wife's income share increases. However, when the wife becomes the primary earner, the husband's non-market labor supply is either stable or decreasing. Across the full relative income range, women remain the main providers of non-market labor, supplying more than twice the number of hours as their husbands, even when they are the breadwinners.

Panels (c) and (d) of Figure 7 replicate these patterns, focusing specifically on hours spent on childcare. The female partner's childcare time declines with her income share, while the male partner's time increases up to the 0.5 threshold but flattens or slightly declines beyond it. While the magnitude of childcare hours is smaller than total non-market work, the behavioral patterns around the threshold are remarkably similar, suggesting that the discontinuity reflects a broader resistance to shifting household responsibilities, even within a narrower domain like childcare. Remarkably, also in this case, the distribution of hours dedicated to childcare by men is always below that of women.

**Figure 6: Relative Income and Distribution of Households According to Hours Spent in Non-Market Work**

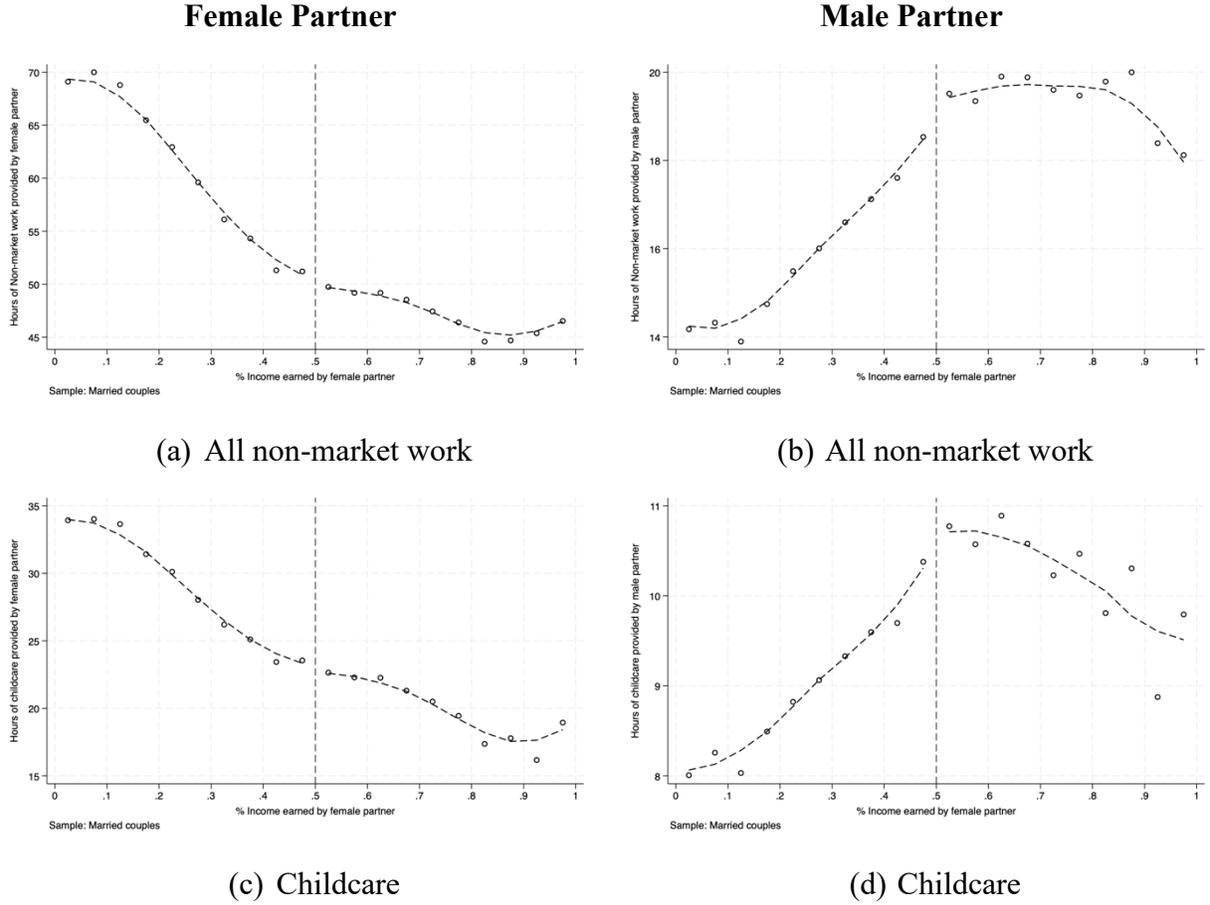

(a) All non-market work

(b) All non-market work

(c) Childcare

(d) Childcare

*Notes:* Data from the 2015 Mexican Intercensal Survey. The sample includes different-sex married couples where both spouses are between 18 and 65 years old report positive labor income. Panels (a) and (b) show average hours spent in all non-market work by the female and male partner, respectively, across the distribution of relative income. Panels (c) and (d) focus specifically on time allocated to childcare. Each point represents the mean within a 0.05-wide bin of relative income, defined as the share of household earnings contributed by the female partner. The vertical line marks the 0.5 threshold, and the curves show LOWESS-smoothed fits.

One advantage of our data is the inclusion of time use information for all adult respondents in the household, which was not present in Bertrand, Kamenica, and Pan (2015). This allows us to formally test the relationship between relative income and the distribution of non-market labor in households. We estimate the following equation:

$$gapNonMarketWork_h = \alpha_0 + \beta_1 wifeEarnsMore_h + \beta_2 relIncome_h \\ + \beta_3 wifeEarnsMore_h \times relIncome_h \\ + \beta_4 \log HHIncome_h + \mu_s + \varepsilon_h \quad (3)$$

Where $gapNonMarketWork_h$ is the number of non-market labor provided by the wife minus the number of hours of non-market labor provided by the husband. $\beta_1$ represents the average difference in the gap in households where the wife is the primary earner; $\beta_2$ captures the evolution of the gap across the relative income distribution for households where the husband is the primary earner; and $\beta_3$ is the change in slope once the wife becomes the primary earner. We also include a control for the log of total household income and state fixed effects $\mu_s$.

**Table 5: Relative Income and Gender Differences in Non-Market Work**

| **Sample:** | All Households | Households without children | Households with children | |
|---|---|---|---|---|
| **Outcome:** | Female-male non-market work hours gap | | | Female-male childcare hours gap |
| | (1) | (2) | (3) | (4) |
| *wifeEarnsMore* | -25.650*** | -17.044*** | -27.597*** | -14.841*** |
| | (0.881) | (1.054) | (1.133) | (0.877) |
| *wifeEarnsMore* X *relIncome* | 0.545*** | 0.356*** | 0.582*** | 0.325*** |
| | (0.015) | (0.018) | (0.019) | (0.015) |
| *relIncome* | -0.646*** | -0.339*** | -0.696*** | -0.424*** |
| | (0.006) | (0.008) | (0.007) | (0.005) |
| Mean of Dep. Var. | 36.057 | 21.252 | 41.391 | 19.756 |
| N | 566,976 | 150,150 | 416,826 | 416,826 |

*Notes:* *** $p<0.01$ ** $p<0.05$ * $p<0.10$. This table presents estimates of the relationship between household relative income and the gender gap in non-market labor hours, measured as the difference between hours supplied by the wife and the husband. The key explanatory variable is a binary indicator for whether the wife is the primary earner, interacted with the continuous measure of her share of household earnings. Column (1) includes all households; columns (2) and (3) stratify by presence of children in the household. Column (4) restricts the sample to households with children and focuses on hours spent on childcare. All regressions include controls for log household income and state fixed effects. Sample restricted to different-sex couples with positive labor earnings and excludes equal-earning couples.

Table 5 shows that, on average, women supply 36 more weekly hours of non-market work than their husbands. This gap is larger for households with children (41 hours) and smaller for households without children (21 hours). For childcare specifically, women supply almost 20 more hours than their husbands. The coefficient on *relIncome* is negative, suggesting that as the woman's share of household income increases, the gender gap in non-market work becomes smaller, consistent with Beckerian expectations of household specialization. Similarly, in households where the wife is the primary earner, the overall gap is smaller, although still positive.

Interestingly, the positive coefficient on the interaction term *wifeEarnsMore* X *relIncome* suggests that once the woman becomes the primary earner, increases in the woman's share of household income are associated with a slower convergence in hours of non-market labor supply.

While this pattern is broadly consistent with Bertrand, Kamenica, and Pan (2015), we do not find evidence that women who outearn their husbands increase their unpaid household labor– potentially to "compensate" for violating gender norms– as they document for the U.S. Instead, the gap declines continuously, although at a slower rate, suggesting a more nuanced norm enforcement.

## 7. Validity and Extensions

To assess the generalizability of our findings, we replicate our main analyses using data from the 2010 Brazilian Census (10% sample) and the 2023 Panamanian Census (full count). We apply consistent sample restrictions and estimation strategies across all three countries. While the Mexican data remain uniquely rich in offering time use measures, the Brazilian and Panamanian censuses still allow us to explore the distribution of relative earnings within households and detect potential discontinuities around the 50% threshold.

In Appendix C we present results for Brazil, where we observe broadly similar patterns to those found in Mexico. For different-sex couples, we again find a sharp drop at the 50% threshold in the wife's share of household income (Figures C1-C2 and Table C1), larger than those discontinuities found in Europe and the US, and comparable to the one in Mexico, thus reinforcing the idea that such discontinuities are not unique to Mexico. These results are consistent with existing research on labor force participation and gender norms in Brazil (Codazzi et al., 2018), which documents the persistence of unequal labor market dynamics within households.

As for Mexico, similar discontinuities are also found over time (Figure C3), in couples with and without children (Figure C4), among unmarried couples (Figure C5), and among female-headed households (Figure C9).

We also detect a significant discontinuity for male same-sex couples, but not for female same-sex couples when focusing on the distribution of the share of total household labor earnings by the non-household head (Figure C6), while we find visual evidence of discontinuities for both male and female same-sex couples when looking at the distribution of the share of earnings by the younger partner (Figure C7). In all cases, the discontinuities are not statistically significant (Table

C3), likely due to sample size, further reinforcing the need for future research to explore intra-household dynamics in same-sex couples by leveraging larger datasets.

The Brazilian data further allow us to explore the potential influence of marriage type. In Latin America, religiosity is often linked to more traditional gender norms (Vaggione & Machado, 2020), which may shape labor market behavior and household dynamics. We compare couples reporting religious marriages (either religious-only or religious and civil) to those with civil marriages only. We find no meaningful differences in the distribution of relative earnings by marriage type (Figure C11). This suggests that formal marriage arrangements rooted in religious practice are not the primary drivers of the observed gaps.

Finally, we replicate the main results using the 2023 Panamanian Census. These analyses are limited to different-sex couples, as the data in the Panamanian Census does not allow for the identification of same-sex partnerships. The distribution of relative earnings in Panama closely mirrors the patterns observed in Mexico and Brazil, including a drop at the 50% threshold (Figures D1-D2 and Table D1). Taken together, these cross-country comparisons underscore the robustness of our findings across Latin America and suggest that these patterns are not idiosyncratic to a single country or institutional context.

## 8. Conclusion

This paper documents a sharp and persistent discontinuity in the distribution of relative income at the 0.5 threshold in Mexico, where the wife begins to earn more than her male partner. The size of the discontinuity– substantially larger than what has been found in high-income countries– suggests strong underlying dynamics shaping intra-household earnings. We show that this pattern is robust across a wide set of specifications and household types, including couples with and without children, households headed by women, and those where the wife is older than the husband. Similar patterns emerge in other Latin American countries, and partially in same-sex couples. We expand the discussion of relative household income beyond Northern Europe and North America, leveraging detailed time-use data and, for the first time, examining income dynamics among same-sex couples. Despite these advances, we are still unable to fully determine whether the observed discontinuity at the 0.5 mark in the relative income distribution is driven primarily by gender norms or by other factors.

Some findings– such as the persistence of the discontinuity even after excluding equal-earning couples, its persistence over time and across countries– suggest that the observed discontinuity is not entirely attributable to co-working arrangements or idiosyncratic confounding variables. However, other results point to a more nuanced picture. For instance, unlike previous US-based research, we do not find consistent evidence that women increase their household work hours when they earn more than their male partners. Additionally, the presence of a discontinuity among some same-sex couples (though not consistently across countries or couple types) suggests that gendered expectations may not be the only explanatory mechanism. These patterns could reflect a combination of internalized norms or other structural factors shaping household specialization. Due to data limitations, however, we are unable to fully disentangle these mechanisms.

Another shortcoming of this study is due to the lack of data on sexual orientation and gender identity. As a result, we cannot distinguish bisexual individuals in either same-sex or different-sex couples, nor can we draw conclusions about couples that include transgender or non-binary individuals. Since these individuals are likely to be gender non-conforming and reject traditional norms, it would be particularly valuable to expand the analyses to these couples and better understand the role of gender norms in driving the discontinuity in relative income.

Collecting longitudinal data from Latin America would allow researchers to examine how income dynamics affect household outcomes such as fertility, division of labor, and divorce– particularly when women earn more than their male partners. Furthermore, linking survey data with administrative and tax records would improve the accuracy of income measures and reduce measurement error. Finally, larger and more inclusive datasets are essential for a deeper understanding of same-sex couples: how they divide household labor, how these patterns evolve over time, and how they may shift after key life events such as becoming parents.

Our findings suggest that policies aimed at reducing the social or economic penalties associated with non-traditional household arrangements, such as women earning more than their partners, could help ease constraints on household decision-making. As already emphasized by Bertrand, Kamenica, and Pan (2015), understanding how gender norms are shaping intra-household decisions such as labor supply, childbearing, household production, marriage and divorce is important for policy-makers, especially in a world of rapidly declining fertility and marriage rates.

**Appendix A. Supplemental Figures and Tables**

**Figure A1: Attitudes towards women outearning their husbands in Brazil and Mexico between 1995 and 2020**

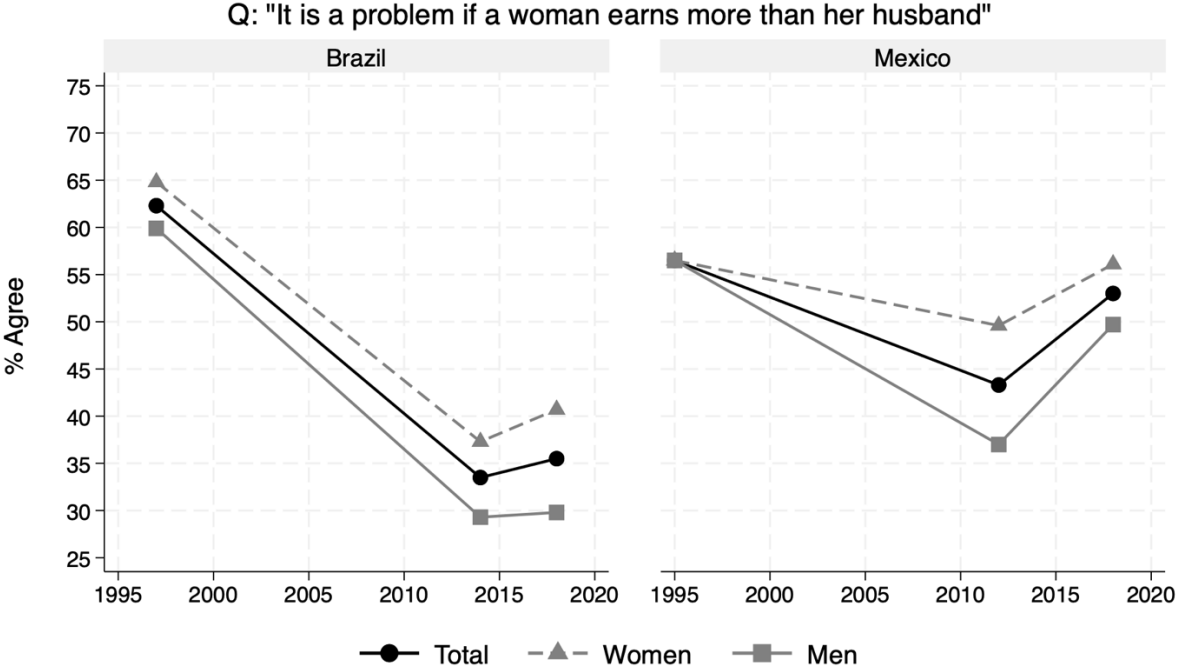

*Notes:* Data from the World Values Survey (WVS), various waves between 1995 and 2020. Respondents were asked whether they agree with the statement: *"It is a problem if a woman earns more than her husband."* The figure reports the share of respondents agreeing, separately by gender, in Brazil and Mexico. Data for Panama are not available.

**Figure A2: Kernel Density of Relative Income - Mexico, 2015**

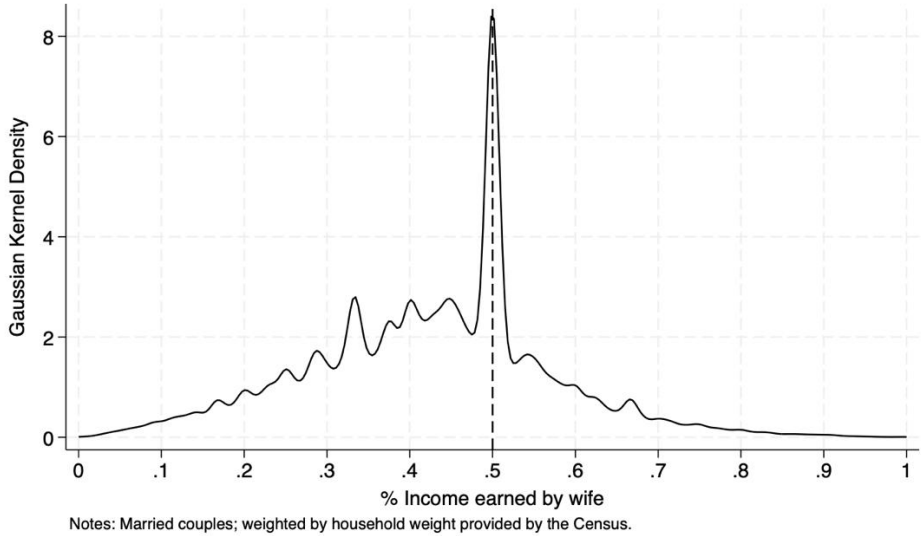

(a) Full sample

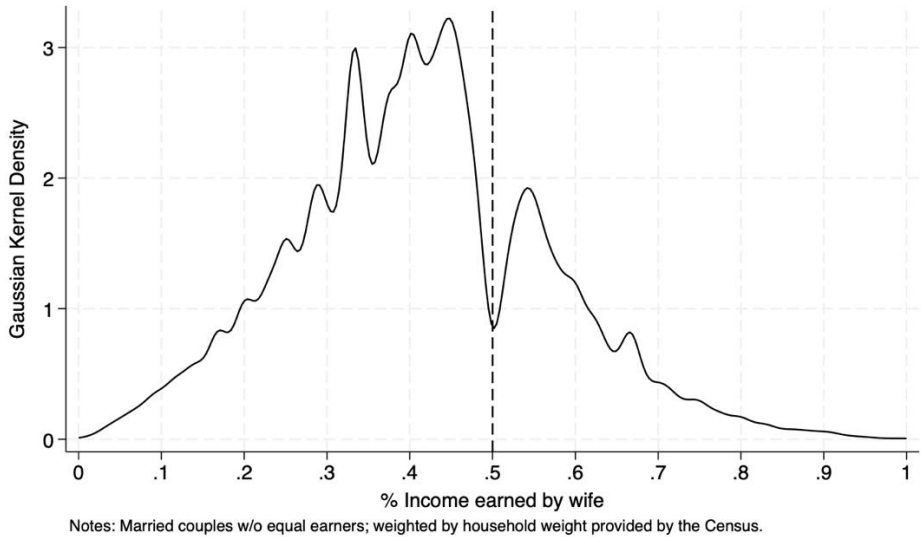

(b) Without equal earners

*Notes:* Data from the 2015 Mexican Intercensal Survey. The sample includes different-sex married couples in which both spouses are between 18 and 65 years old and report positive labor income. Densities are estimated using a Gaussian kernel and household weights provided by the Census. Panel (a) includes all couples; Panel (b) excludes equal earners. The vertical line at 0.5 marks the point at which the wife earns more than her husband.

**Figure A3: Distribution of Relative Income After Adding Random Noise to Individual Income**

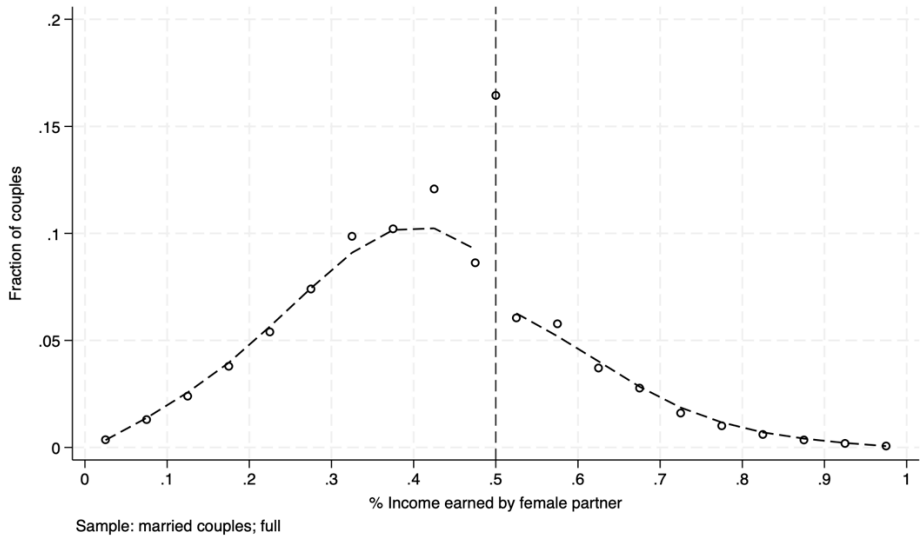

(a) Full sample

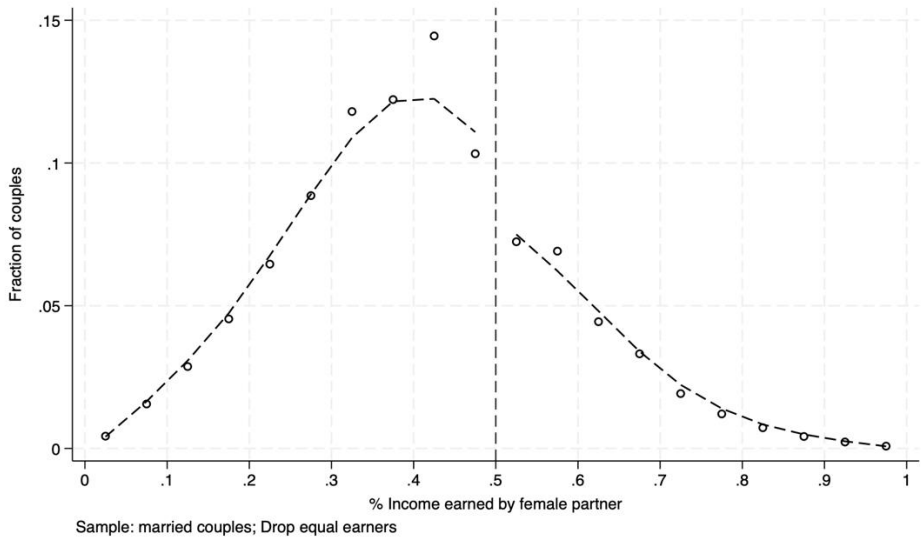

(b) Without equal earners

*Notes:* Data from the 2015 Mexican Intercensal Survey. The sample includes different-sex married couples in which both spouses are between 18 and 65 years old and report positive labor income. We add random noise equal to ±1% of each individual's labor income before computing relative earnings. Panel (a) shows the full sample; Panel (b) excludes couples with exactly equal earnings. Each point represents the share of couples within a 0.05-wide bin of relative income. The vertical line marks the 0.5 threshold.

**Figure A4: Distribution of the share of total household labor earnings by the wife in Mexico in 2000, 2010, and 2020**

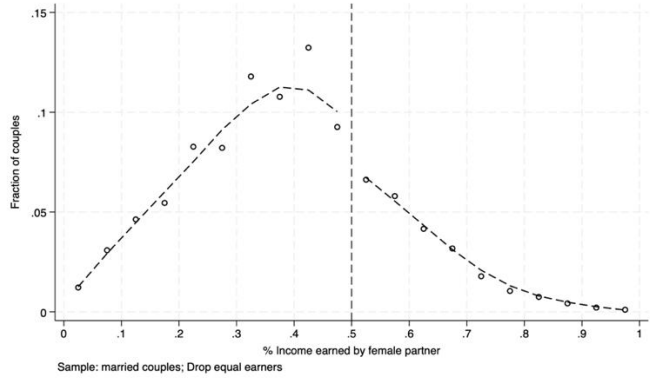

(a) Mexico (2000)

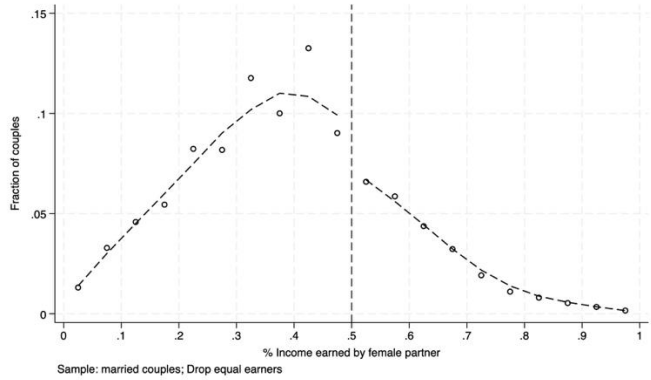

(b) Mexico (2010)

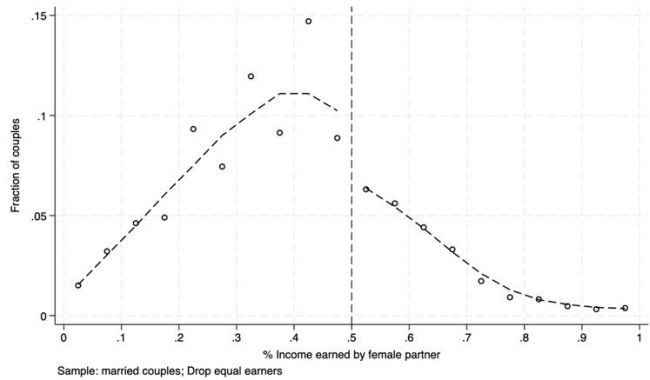

(c) Mexico (2020)

*Notes:* Data from the 2000, 2010, and 2020 Mexican Censuses. The sample includes different-sex married couples in which both spouses are between 18 and 65 years old and report positive labor income. Couples with identical reported incomes are excluded. Each point represents the share of couples within a 0.05-wide bin of relative income earned by the wife. The vertical line marks the 0.5 threshold, and the curve shows a LOWESS-smoothed fit to the distribution.

**Figure A5: Distribution of the relative income earned by the wife under random coupling**

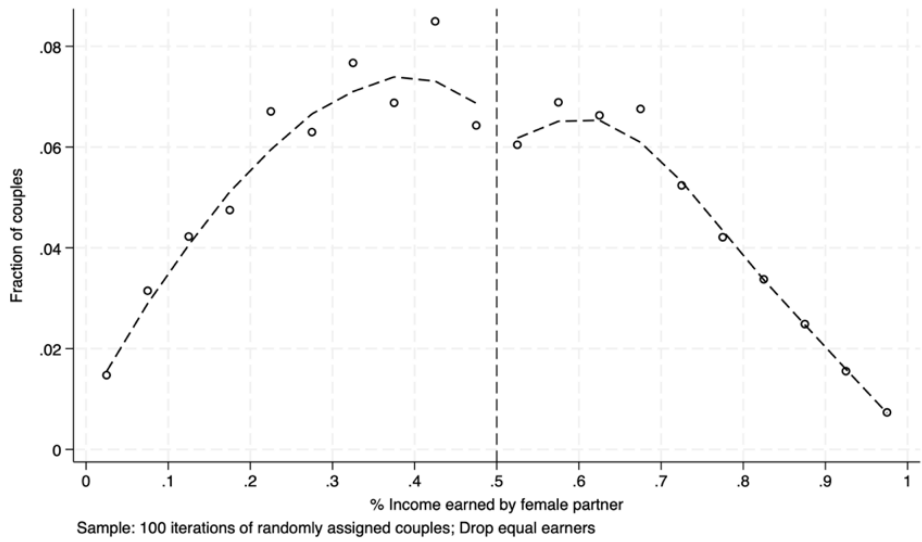

*Notes:* Data from the 2015 Mexican Intercensal Survey. Each point represents the share of couples within a 0.05-wide bin of the wife's income share, based on 100 iterations of random matching using different-sex couples. In each iteration, we randomly pair men and women drawn from the same sample as in Figure 1, excluding equal-earning couples. The vertical line marks the 0.5 threshold, and the curve shows a LOWESS-smoothed fit to the simulated distribution.

**Figure A6: Relative Income Earned by Younger Partner and Non-Household Head**

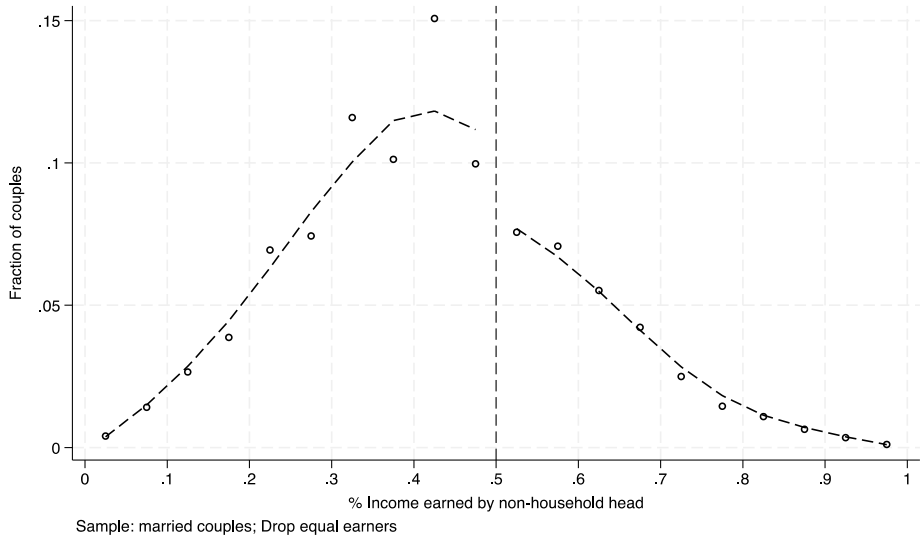

(a) Relative income earned by the non-household head

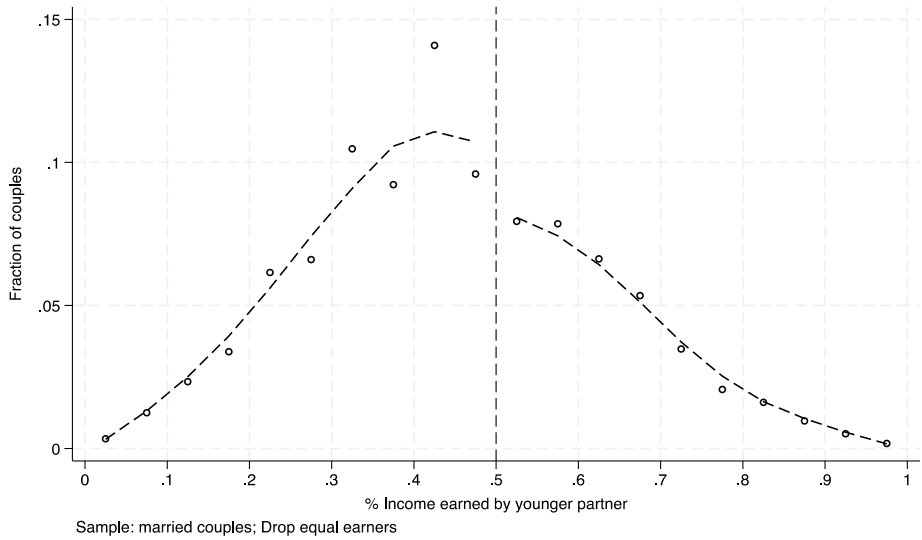

(b) Relative income earned by the younger partner

*Notes:* Data from the 2015 Mexican Intercensal Survey. The sample includes different-sex married couples in which both spouses are between 18 and 65 years old and report positive labor income. Couples with identical reported incomes are excluded. Each point represents the share of couples within a 0.05-wide bin of relative income earned by (a) the non-household head and (b) the younger partner. The vertical line marks the 0.5 threshold, and the curve shows a LOWESS-smoothed fit to the distribution.

**Figure A7: Distribution of households according to relative income in households where the wife is older than her husband**

**Mexico**

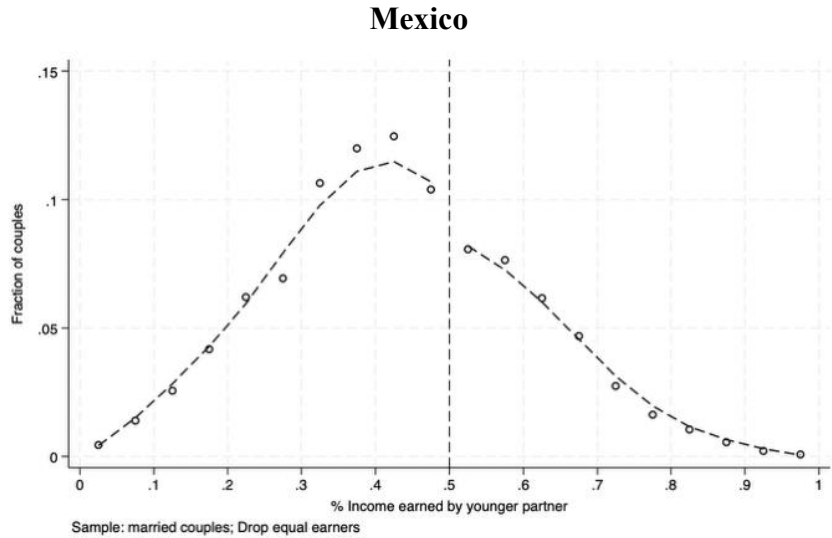

(a) Wife is older by >3 years

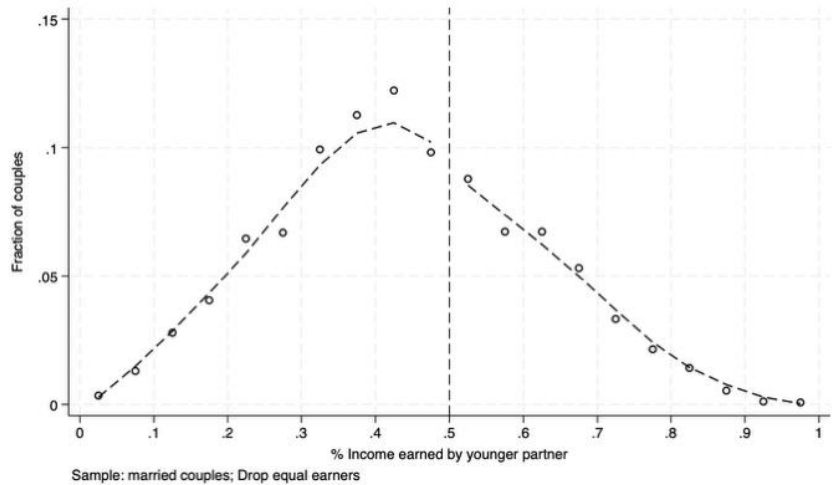

(b) Wife is older by >10 years

*Notes:* Data from the 2015 Mexican Intercensal Survey. The sample includes different-sex married couples in which both spouses are between 18 and 65 years old and report positive labor income. Couples with identical reported incomes are excluded. Each point represents the share of couples within a 0.05-wide bin of relative income earned by the younger partner. Panel (a) restricts to couples where the wife is more than 3 years older than the husband; Panel (b) restricts to those where she is more than 10 years older. The vertical line marks the 0.5 threshold, and the curve shows a LOWESS-smoothed fit to the distribution.

**Figure A8: Distribution of the share of total household labor earnings by the younger partner in same-sex couples in Mexico**

**Mexico**

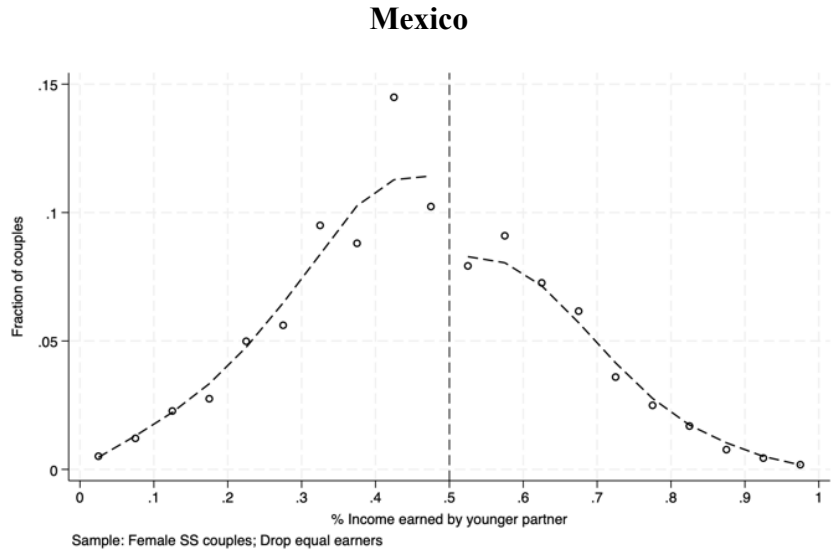

(a) Female same-sex couples

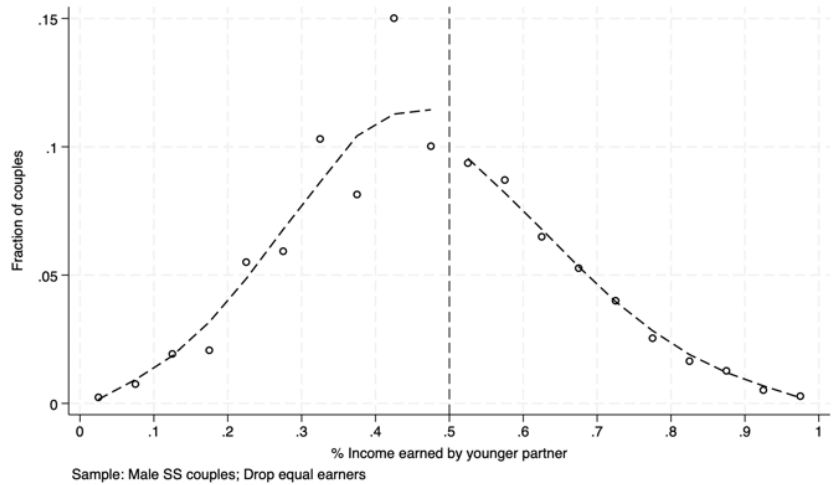

(b) Male same-sex couples

*Notes:* Data from the 2015 Mexican Intercensal Survey. The sample includes same-sex couples in which both partners are between 18 and 65 years old and report positive labor income. Couples with identical reported incomes are excluded. Each point represents the share of couples within a 0.05-wide bin of relative income earned by the younger partner. Panel (a) includes female same-sex couples; Panel (b) includes male same-sex couples. The vertical line marks the 0.5 threshold, and the curve shows a LOWESS-smoothed fit to the distribution.

**Figure A9: Labor Force Participation Rate According to Median Income in Demographic Group**

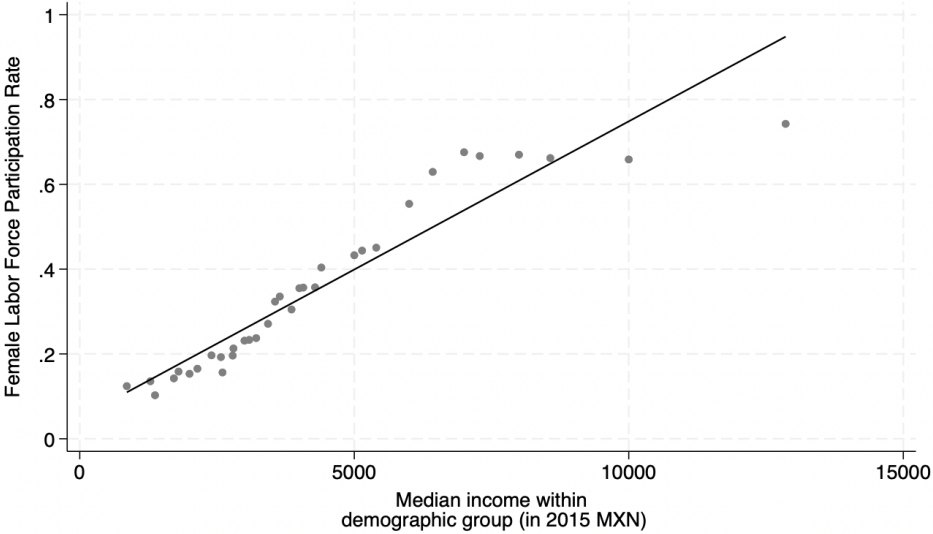

*Notes:* Each dot represents a demographic group defined by race, five-year age bins, education, and state of residence in the 2015 Mexican Intercensal Survey. The horizontal axis shows the median monthly labor income of individuals in each group (in 2015 Mexican pesos), while the vertical axis shows the corresponding female labor force participation rate. The line displays the best linear fit.

**Figure A10: Relative Income and Distribution of Households According to Hours Spent in Non-Market Work – Including Equal earners**

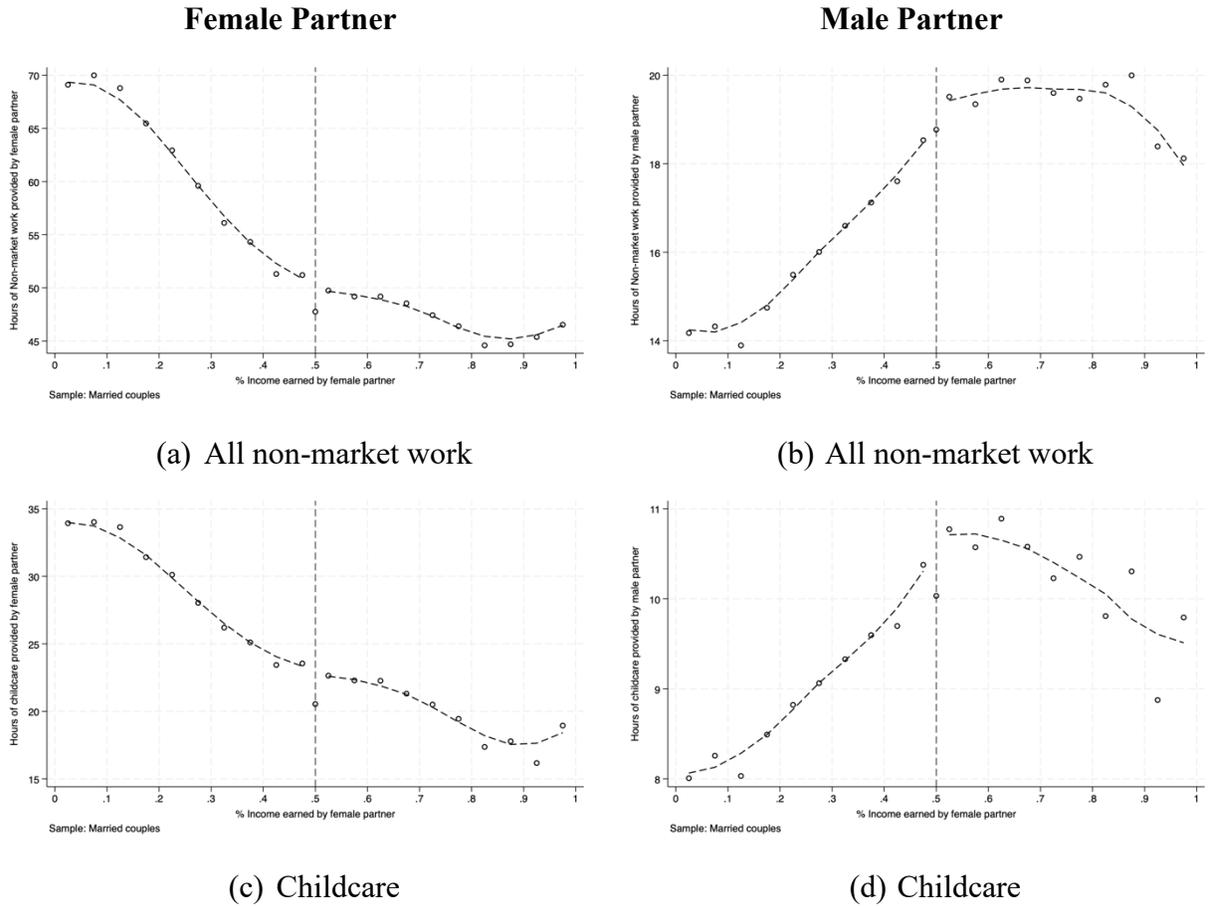

(a) All non-market work

(b) All non-market work

(c) Childcare

(d) Childcare

*Notes:* Each panel plots hours spent in non-market work (top row) or childcare (bottom row) by the female (left column) or male (right column) partner, against the share of household labor income earned by the wife. Unlike the baseline analysis, these graphs include couples where both partners earn exactly the same amount. Sample includes married couples in Mexico in 2015.

**Tables**

**Table A1: Differences between workers in the 0.5 mass point and surrounding areas**

|  | Relative Income Range: | |
|---|---|---|
|  | = .5 | $\in (.4, .6) \setminus \{.5\}$ |
|  | (3) | (4) |
| *Occupation* | | |
| Agricultural workers | 0.083 | 0.058 |
| Teachers | 0.109 | 0.081 |
| Retail workers | 0.219 | 0.148 |
| | | |
| *Industry* | | |
| Agricultural industry | 0.092 | 0.069 |
| Wholesale/retail | 0.238 | 0.169 |
| Hotels/restaurants | 0.103 | 0.078 |
| Education | 0.137 | 0.118 |
| | | |
| Income < Min. Wage | 0.070 | 0.039 |
| Income < 1.5x Min. Wage | 0.202 | 0.135 |
| Self-employed | 0.377 | 0.195 |

*Notes:* This table presents summary statistics for individuals in couples where the wife earns exactly 50% of household income (column 3) and those in nearby ranges of the income distribution, excluding equal earners (column 4). The sample includes different-sex couples from the 2015 Mexican Intercensal Survey. Values correspond to the share of individuals in each category.

**Table A2: McCrary Test for discontinuity in the distribution of the woman's share of total household labor income with alternative cutoffs - Different-Sex Married Partners**

|  | Full Sample | Excluding employers/self employed | Excluding couples in same occupation and industry | Excluding employers/self emp./same occ. and industry |
|---|---|---|---|---|
| | *All samples below exclude equal-earning couples* | | | |
| | (1) | (2) | (3) | (4) |
| **Panel A: 0.5** | | | | |
| log distance at 0.5 | -0.220*** | -0.248*** | -0.176*** | -0.200*** |
| | (0.010) | (0.012) | (0.010) | (0.013) |
| **Panel B: 0.49999** | | | | |
| log distance at 0.49999 | -0.224*** | -0.250*** | -0.179*** | -0.200*** |
| | (0.010) | (0.012) | (0.010) | (0.013) |
| N | 362,030 | 211,517 | 331,978 | 191,820 |

*Notes:* *** $p<0.01$ ** $p<0.05$ * $p<0.10$. McCrary test for discontinuity in the distribution of the woman's share of total household labor income at alternative thresholds (0.5 and 0.49999). Sample includes different-sex married couples from the 2015 Mexican Intercensal Survey, excluding equal-earning couples.

**Table A3: McCrary Test for discontinuity in the distribution of the woman's share of total household labor income – By Parenting Status**

| | Full Sample | Excluding employers/self employed | Excluding couples in same occupation and industry | Excluding employers/self emp./same occ. and industry |
|---|---|---|---|---|
| | *All samples below exclude equal-earning couples* | | | |
| | (1) | (2) | (3) | (4) |
| **Panel A: Households with children** | | | | |
| log distance at 0.50001 | -0.238*** | -0.267*** | -0.193*** | -0.215*** |
| | (0.011) | (0.014) | (0.012) | (0.015) |
| N | 269,205 | 162,588 | 247,050 | 147,322 |
| **Panel A: Households without children** | | | | |
| log distance at 0.50001 | -0.156*** | -0.168*** | -0.123*** | -0.122*** |
| | (0.020) | (0.025) | (0.021) | (0.027) |
| N | 92,825 | 87,579 | 84,928 | 44,498 |

*Notes:* *** $p<0.01$ ** $p<0.05$ * $p<0.10$. McCrary test estimates by parenting status for different-sex married couples in the 2015 Mexican Intercensal Survey, excluding equal-earning couples.

**Table A4: McCrary Test for discontinuity in the distribution of the woman's share of total household labor income - Different-Sex Unmarried Partners**

|  | \multicolumn{4}{c}{*All samples below exclude equal-earning couples*} | | | |
| --- | --- | --- | --- | --- |
|  | Full Sample | Excluding employers/self employed | Excluding couples in same occupation and industry | Excluding employers/self emp./same occ. and industry |
|  | (1) | (2) | (3) | (4) |
| log distance at 0.50001 | -0.247*** | -0.289*** | -0.205*** | -0.236*** |
|  | (0.015) | (0.018) | (0.016) | (0.020) |
| N | 141,680 | 87,579 | 129,648 | 79,120 |

*Notes:* *** $p<0.01$ ** $p<0.05$ * $p<0.10$. Each cell reports the estimated discontinuity in the density of the woman's share of total household labor income using the McCrary (2008) test. The dependent variable is the log difference in the density just above versus just below the threshold. All samples exclude couples with exactly equal earnings. Panel A restricts the sample to households with children; Panel B to those without. Column (1) includes the full sample of different-sex married couples. Column (2) excludes self-employed and employer couples. Column (3) excludes couples where both spouses work in the same occupation and industry. Column (4) excludes both self-employed/employers and couples in the same occupation and industry. Standard errors are in parentheses.

**Table A5: Summary Statistics for 2000 Mexican Sample**

| Couple type: | Different-sex, married | | Different-sex, cohabiting | |
|---|---|---|---|---|
| Sex: | Women | Men | Women | Men |
| ***Individual-level Variables*** | | | | |
| Age | 36.066 | 38.801 | 33.712 | 36.597 |
| | (8.577) | (9.337) | (8.756) | (10.030) |
| High school degree | 0.426 | 0.414 | 0.223 | 0.227 |
| | (0.494) | (0.492) | (0.416) | (0.419) |
| College degree | 0.182 | 0.219 | 0.072 | 0.091 |
| | (0.386) | (0.413) | (0.259) | (0.287) |
| Income | 3324.29 | 5440.48 | 2422.26 | 3728.65 |
| | (9942.09) | (14002.17) | (9431.17) | (11417.25) |
| Self-employed | 0.290 | 0.266 | 0.271 | 0.254 |
| | (0.454) | (0.442) | (0.444) | (0.435) |
| Household Head | 0.030 | 0.970 | 0.089 | 0.910 |
| | (0.290) | (0.442) | (0.285) | (0.285) |
| Older Partner | 0.173 | 0.701 | 0.270 | 0.636 |
| | (0.378) | (0.457) | (0.444) | (0.481) |
| ***Household-level variables*** | | | | |
| Relative income | 0.387 | | 0.388 | |
| | (0.172) | | (0.167) | |
| Both self-employed | 0.113 | | 0.107 | |
| | (0.317) | | (0.309) | |
| Same occupation and industry | 0.141 | | 0.146 | |
| | (0.141) | | (0.353) | |
| Any children in HH | 0.689 | | 0.709 | |
| | (0.463) | | (0.454) | |
| Any children <5 in HH | 0.347 | | 0.380 | |
| | (0.476) | | (0.485) | |
| Households | 191,040 | | 41,326 | |

*Notes*: Summary statistics for different-sex married and cohabiting couples in the 2000 Mexican Census, excluding equal-earning couples. Means and standard deviations (in parentheses) are shown for individual- and household-level variables, separately by sex. Household-level variables refer to characteristics of the couple or household. "Older Partner" indicates the share of individuals who are older than their partner. "Relative income" refers to the woman's share of total household labor income.

**Table A6: McCrary test for discontinuities in the distribution of the non-household head/younger partner's share of total labor earnings - Same-Sex Partners**

|  | log distance at 0.50001 | |
| --- | --- | --- |
|  | Female S.S. households | Male S.S. households |
|  | (1) | (2) |
| **Panel A: Non-household head partner** | | |
| All dual-earning couples | -0.250** | 0.183 |
|  | (0.124) | (0.141) |
| N | 2,726 | 2,125 |
| **Panel B: Younger Partner** | | |
| All dual-earning couples | -0.230* | -0.008 |
|  | (0.125) | (0.140) |
| N | 2,726 | 2,125 |

*Notes:* *** $p<0.01$ ** $p<0.05$ * $p<0.10$. Each cell reports the estimated discontinuity in the density of the share of total labor earnings contributed by the non-household head or younger partner in same-sex couples, using the McCrary (2008) test. The dependent variable is the log difference in the density just above versus just below the 50% threshold. All samples are restricted to dual-earning same-sex couples. Panel A defines the reference partner as the household head, and Panel B defines the reference partner as the older partner. Standard errors are in parentheses.

**Table A7: Earnings Potential and Realized Wife's Income - Using the median of the potential income distribution**

|  | (1) | (2) | (3) | (4) |
|---|---|---|---|---|
| *PrWifeEarnsMore* | -0.064*** | -0.024** | -0.012 | -0.013 |
|  | (0.011) | (0.011) | (0.011) | (0.008) |
|  |  |  |  |  |
| Mean of *incomeGap* | 0.129 | 0.129 | 0.129 | 0.129 |
| $N$ | 616,556 | 616,556 | 616,556 | 616,556 |
| $R^2$ | 0.158 | 0.167 | 0.170 | 0.170 |
|  |  |  |  |  |
| *Additional controls:* |  |  |  |  |
| Cubic in *logHusbIncome* |  | X | X | X |
| Children in household |  |  | X | X |
| Children under 5 in household |  |  |  | X |

*Notes:* *** $p<0.01$ ** $p<0.05$ * $p<0.10$. Each cell reports estimates from a regression of the income gap on the probability that the wife would outearn her husband. The income gap is defined using the median of the potential income distribution for women with similar demographic characteristics. All specifications include fixed effects for the wife's demographic group. Additional controls are included as indicated. Standard errors are in parentheses.

**Appendix B. Data Sources and Construction**

**B1. Variable description**

*Sex* reports whether the person was male or female. Our data do not allow us to distinguish between sex and gender. In all of our datasets, there are no flags indicating whether sex was imputed.

Each household contains an individual designated as the household head. All other household members report their relationship to the household head according to a pre-defined set of categories. Couples are identified when an individual reports being the spouse or partner of the household head. We define same-sex couples where both partners in a household (household head and spouse/partner) report having the same sex.

*Age* reports the respondent's age in years at the time of the interview.

*Ethnicity and race.* Ethnicity is a multidimensional concept that can be measured using a diverse set of approaches, including ethnic ancestry or origin, ethnic identity, cultural origins, nationality, race, color, minority status, language, religion, or various combinations of them. The countries in our sample asked individuals to self-identify phrasing the question including some of the concepts previously listed. In Brazil, individuals could choose between *white, yellow, brown, black, or indigenous*. For those who do not self-identify as *indigenous,* a supplemental yes/no question was presented to ask whether they consider themselves indigenous. In Mexico and Panama, individuals were asked yes/no questions about belonging to any indigenous people or to the African descendant community.

*Education*. We use discrete classifications which emphasize the highest-level completed by the respondent. For Brazil, the options are *less than primary completed, primary completed, secondary completed,* and *university completed.* For Mexico, the options are *no education, preschool, primary completed, secondary completed, high school (general), high school (technical), vocational degree with primary completed, vocational degree with secondary completed, vocational degree with high school completed, university degree, masters degree,* and *doctoral degree.* For Panama, options include *primary school incomplete, primary school complete, secondary school incomplete,* and *1$^{st}$ cycle of secondary complete, 2$^{nd}$ cycle of secondary complete.*

*Income* is a continuous variable that indicates monthly labor income in national currency. In Brazil, the variable refers to the respondent's earnings for their main job in July 2010 and does not include retirement, pension, social programs, transfers, and other sources. In Mexico, the labor income includes earnings for the respondent's main job. In Panama, income is measured in U.S. Dollars and refers to monthly labor income.

*Work Position.* In Brazil and Mexico, we define a couple in the same occupation and position if both individuals report working in the same occupation and in the same industry. The Panamanian census did not collect information on occupation. Thus, we classify an individual's work *position* according to their response of the following statement: In your main job, [RESPONDENT] was: *employee/laborer*, *day laborer/farmhand, paid assistant, employer, self-employed worker,* and *unpaid worker*.

*Marital status.* In Brazil, respondents were given the following options: *married, separated, divorced, widowed,* and *single*. Respondents who reported being married were asked a subsequent question about the nature of their marital union. The possible categories were *civil and religious marriage, only civil marriage, only religious marriage,* and *consensual union*. In Mexico, respondents had the following options: *live with partner in a free union, separated, divorced, widowed, married, single*. In Panama, options include *single, formal/informal union, divorced/separated,* and *widowed*. For our samples of couples, we restrict to those reporting being married or in a domestic partnership (*consensual union* in Brazil or *free union* in Mexico).

*Household Head.* In Brazil, in 2010, the household head is defined as the person who is recognized as such by the other household residents. The census enumerator was instructed to fill out who was recognized as the household head. In Mexico, the household head portion of the questionnaire reads as follows: "Please tell me the names of everyone who normally lives in this home, including young children and the elderly. Also include any domestic staff who sleeps here. Start with the head of the household." The Panamanian census does not give any instructions on how to define household head. Instead, respondents are just asked to fill out the individual information sheet starting by the household head.

**Figure B1: Household Composition Questionnaires**

(a) Household Questionnaire in the 2010 Brazilian Census

(b) Household Questionnaire in the 2015 Mexican Intercensal Survey

| Persona N.° | 1. Nombres y apellidos (La primera persona de la lista debe ser el jefe o la jefa del hogar) | 2. Sexo | | 3. Edad |
| --- | --- | --- | --- | --- |
| | | Hombre | Mujer | Años cumplidos |
| 01 | | ◯ 1 | ◯ 2 | ☐☐☐ |
| 02 | | ◯ 1 | ◯ 2 | ☐☐☐ |

**IV. LISTA DE OCUPANTES DEL HOGAR**

A CONTINUACIÓN, ANOTAREMOS A LOS RESIDENTES HABITUALES DEL HOGAR. TENGA EN CUENTA QUE EL RESIDENTE HABITUAL ES LA PERSONA QUE VIVE HABITUALMENTE O DUERME LA MAYOR PARTE DEL TIEMPO EN LA VIVIENDA, ASÍ COMO AQUELLAS QUE TIENEN LA INTENCIÓN DE RESIDIR AQUÍ.

DÍGAME LOS NOMBRES Y APELLIDOS DE TODAS LAS PERSONAS QUE RESIDEN HABITUALMENTE EN ESTE HOGAR, EMPEZANDO CON EL JEFE O JEFA Y CONTINÚE EN EL SIGUIENTE ORDEN: CÓNYUGE, LOS HIJOS(AS) SOLTEROS(AS) DE MAYOR A MENOR, LOS HIJOS(AS) CASADOS(AS) CON SUS CÓNYUGES E HIJOS, OTROS PARIENTES, LOS NO PARIENTES Y LOS MIEMBROS DEL SERVICIO DOMÉSTICO QUE RESIDEN HABITUALMENTE EN EL HOGAR.

(c) Household Questionnaire in the 2023 Panamanian Census

**B2. Time Use data in Mexico**

In 2015, the Mexican Intercensal Survey included questions regarding time use in household-related activities. All individuals over 12 years old were asked to estimate the number of hours, between 0 and 140, spent on the following categories:

| | Categories | Possible Range |
|---|---|---|
| **In the past week, without getting paid, how many hours have you dedicated to:** | Attending to individuals with special needs (help with food, movement, medications, etc.) | 0-140 hours |
| | Attending to infirm individuals who require special needs (help with food, movement, medications, etc.) | 0-140 hours |
| | Attending to a healthy child under 6 years old (children, grandchildren, nieces/nephews; help with food, dressing, take them to school, etc.) | 0-140 hours |
| | Attending to a healthy child between 6-14 years old (children, grandchildren, nieces/nephews; help with food, dressing, take them to school, etc.) | 0-140 hours |
| | Attending to an individual over 60 years old who requires continuous care? (Parents, grandparents; help with food, doctor appointments, dressing, etc.) | 0-140 hours |
| | Preparing or serving food to your family | 0-140 hours |
| | Cleaning your home, washing/ironing your family's clothes | 0-140 hours |
| | Shopping for food or cleaning supplies | 0-140 hours |

As individuals can report up to 140 hours spent on each activity, it is possible that the weekly number of hours spent in non-market work exceeds 168. Mexico's statistical office (INEGI) does not recode the number of hours for cases in which this occurs.

**Appendix C: Results for Brazil**

We present evidence from the 2010 Brazilian Census, using the same approach as in the Mexican analysis. The results reveal similar discontinuities– slightly larger in magnitude– in the distribution of the woman's share of household labor income. This suggests that the behavioral patterns we document are not unique to Mexico and may reflect broader dynamics across Latin America.

We further examine whether these discontinuities vary by marital status. Discontinuities are larger for unmarried couples, consistent with prior research on the rise of cohabitation in the region during the 21st century (Esteve et al., 2012). Uniquely, the Brazilian census distinguishes between civil marriages, religious marriages, and unions that are both civil and religious. While this allows us to explore heterogeneity by union type– potentially capturing more traditional views on gender roles– we caution that these categories may have been inconsistently understood by respondents. In Appendix Figure C11, we find no meaningful differences in discontinuities between couples who report any religious component to their union and those who do not.

We also include results for same-sex couples. While estimates are less precise, we find evidence of a discontinuity in the distribution of relative income for male same-sex couples, but not for female same-sex couples.

Taken together, the Brazilian findings reinforce the idea that the observed discontinuities reflect structural features of gender and intra-household bargaining that extend beyond any single national context.

This section is organized as follows. Figure C1 show the Gaussian Kernel Density of relative income (Figure A2 for Mexico). Figures C2 and C3 show the relative income distribution in Brazil for 2010 and 2000, respectively (Figures 1 and A4 for Mexico). Figure C4 shows the heterogeneous distributions according to parenting status (Figure 2, Panels (a) and (b)). Figure C5 shows the distribution for unmarried, cohabiting couples in Brazil (equivalent to Figure 2, Panel (d)). Figures C6 and C7 show the distribution of relative income in same-sex households according to household headship and age (Figures 5 and A8). Figure C8 shows the counterfactual distribution of relative income in Brazil under random sorting (Figure A5). Figure C9 shows the distribution for households where the wife is the household head (Figure 3). Figure C10 shows the distribution for households where the wife is older than her husband (Figures 4 and A7). Figure C11 shows the distribution according to union type (religious and non-religious).

**Figure C1: Kernel Density of Relative Income – Brazil, 2010**

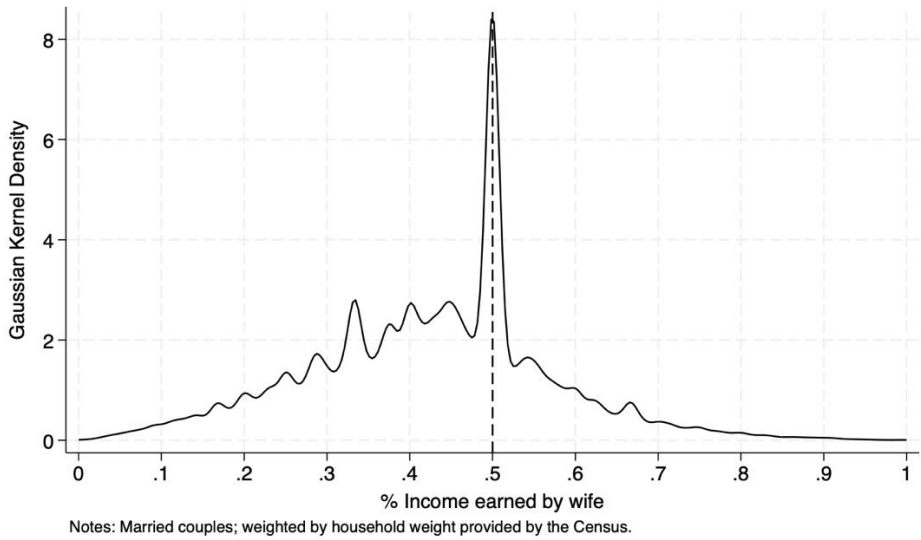

(a) Full sample

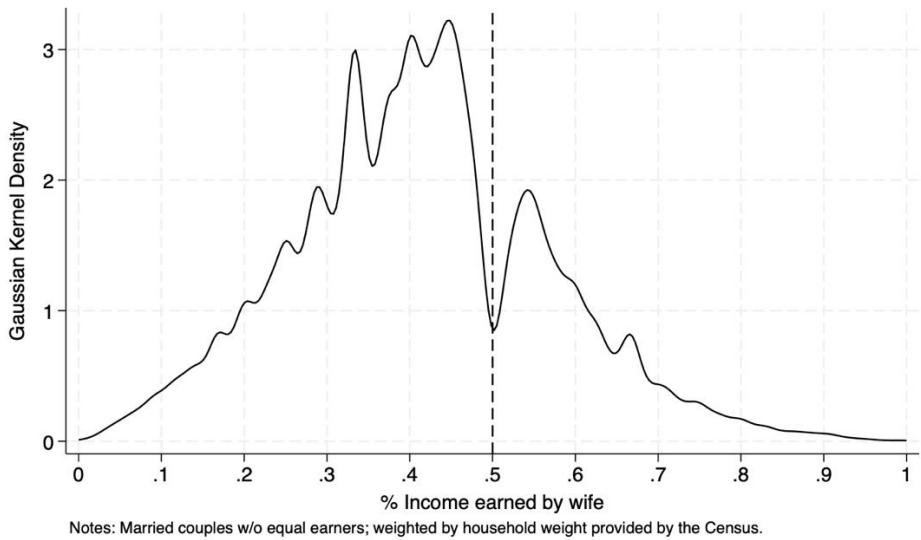

(b) Without equal earners

*Notes:* Sample includes married different-sex couples in the 2010 Brazilian Census (10% sample). Panel (a) shows the distribution of the wife's share of household labor income, using Gaussian kernel density and household sampling weights. Panel (b) drops couples with equal earnings.

**Figure C2: Distribution of the share of total household labor earnings by the wife in Brazil**

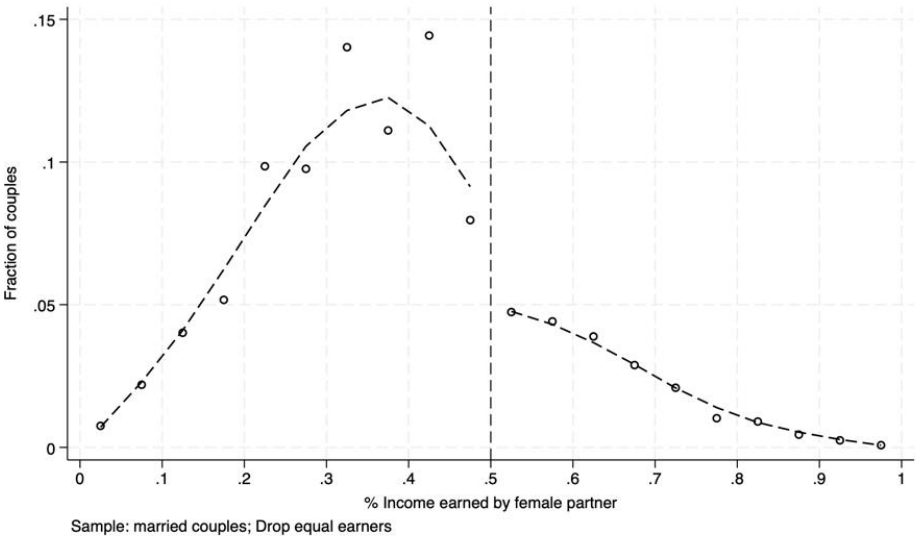

(a) Brazil

*Notes:* This figure plots the distribution of the wife's share of total household labor income using data from the 2010 Brazilian Census (10% sample). The sample includes married different-sex couples and excludes equal earners. The vertical line marks the 50% threshold in relative income.

**Figure C3: Distribution of the share of total household labor earnings by the wife in Brazil in 2000**

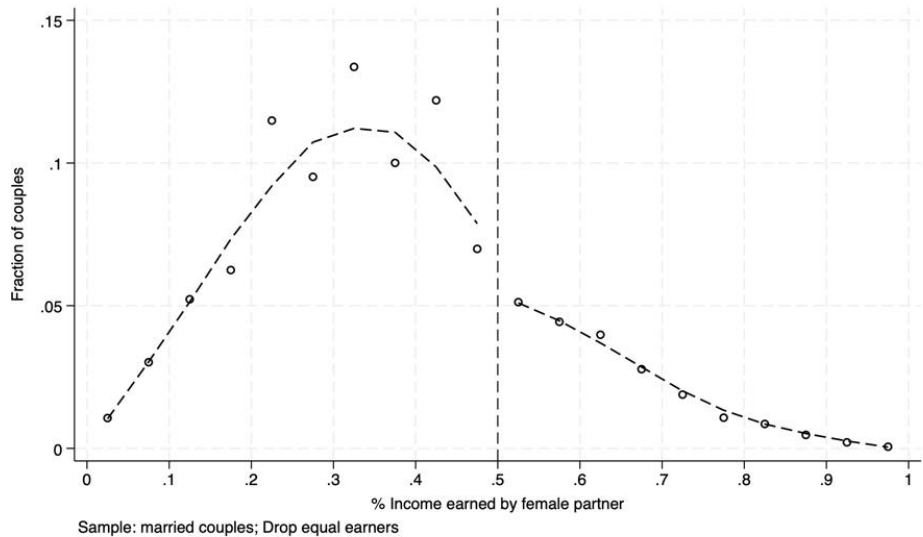

(a) Brazil (2000)

*Notes:* This figure plots the distribution of the wife's share of total household labor income using data from the 2000 Brazilian Census (10% sample). The sample includes married different-sex couples and excludes equal earners. The vertical line marks the 50% threshold in relative income.

**Figure C4: Distribution of couples of relative earnings of female partner according to parenting status in Brazil**

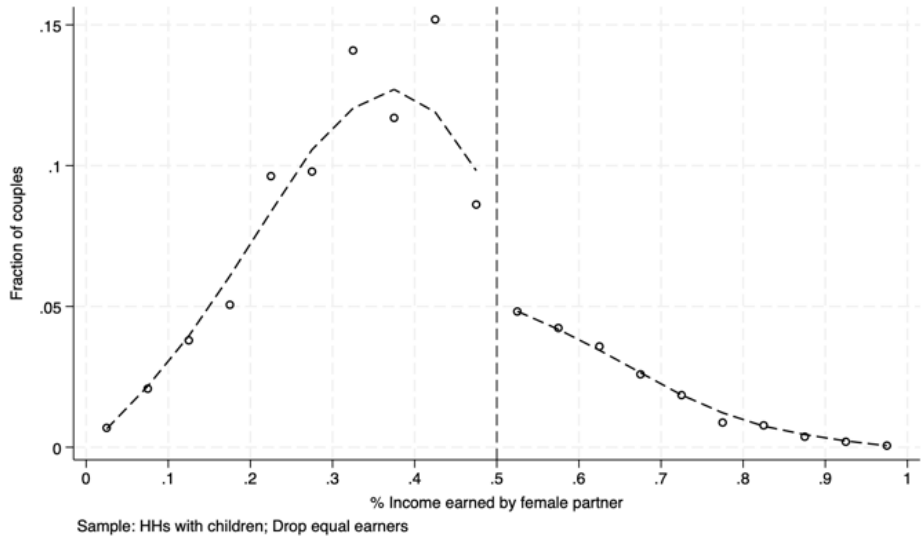

(a) With children

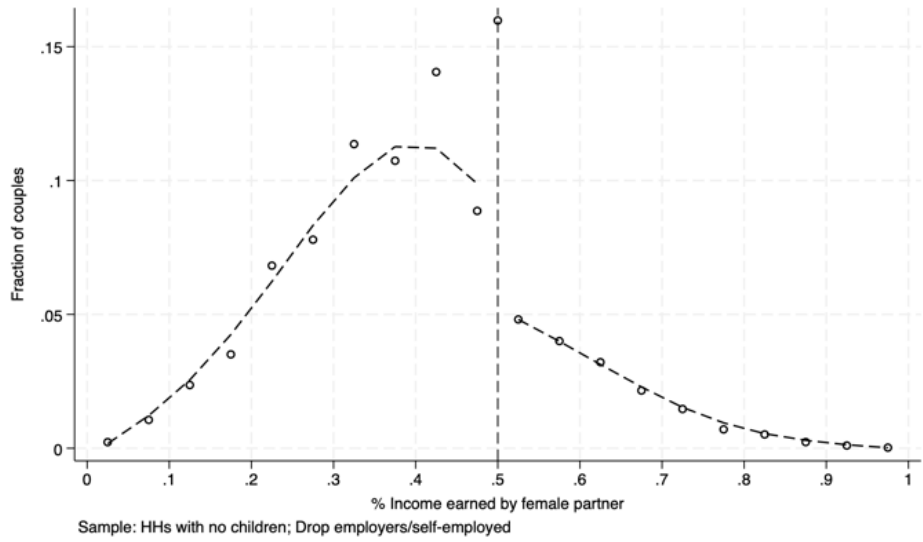

(b) Without children

*Notes:* This figure uses data from the 2010 Brazilian Census (10% sample) to show the distribution of the female partner's share of household labor income by parenting status. Panel (a) restricts the sample to households with children, while panel (b) includes only households without children and excludes employers and self-employed individuals. Equal earners are dropped in both panels. The vertical line marks the 50% income share threshold.

**Figure C5: Discontinuities for Unmarried, Cohabiting Couples in Brazil**

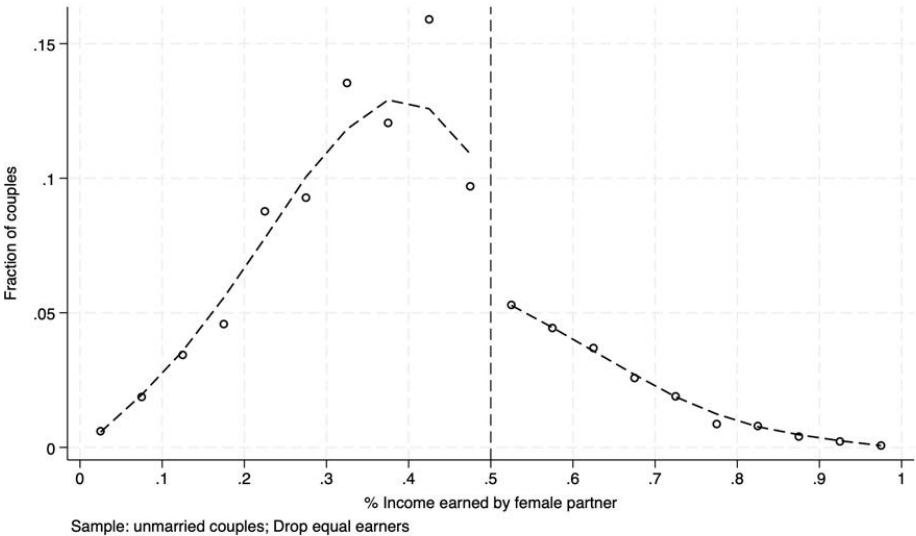

*Notes:* This figure uses data from the 2010 Brazilian Census (10% sample) to examine the distribution of relative earnings among unmarried, cohabiting couples. The sample excludes equal earners.

**Figure C6: Distribution of the share of total household labor earnings by the non-household head in same-sex couples in Brazil**

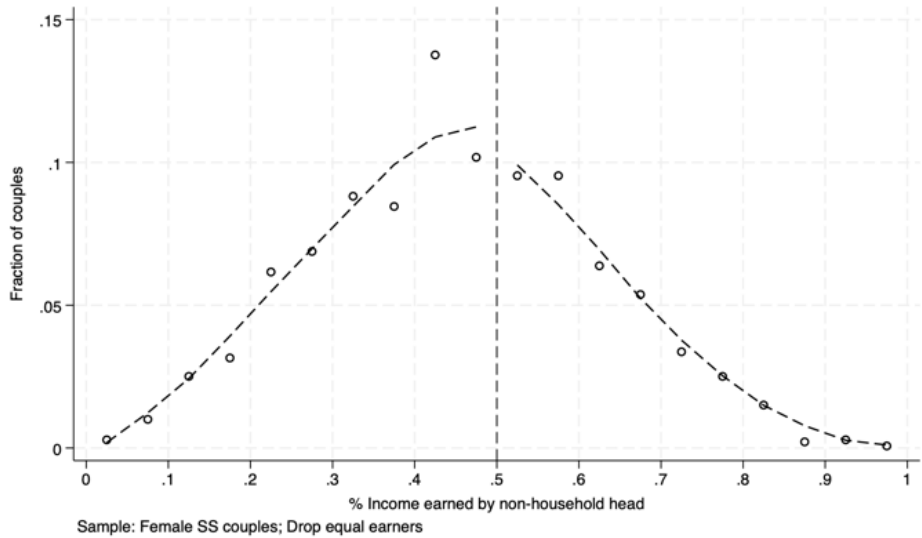

(a) Female same-sex couples

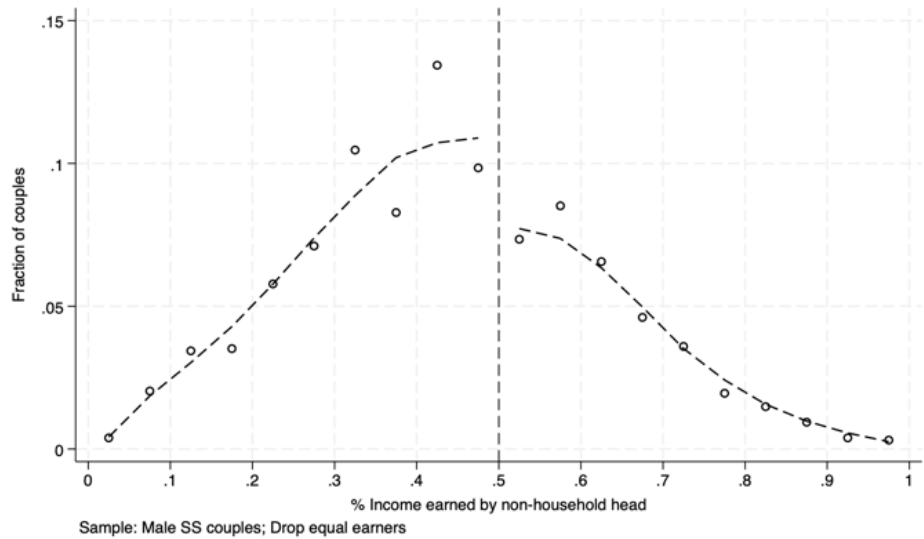

(b) Male same-sex couples

*Notes:* This figure uses data from the 2010 Brazilian Census (10% sample). It shows the distribution of the share of total household labor earnings earned by the non-household head in same-sex couples. Equal earners are excluded. Panel (a) displays the distribution for female same-sex couples, and Panel (b) for male same-sex couples.

**Figure C7: Distribution of the share of total household labor earnings by the younger partner in same-sex couples**

**Brazil**

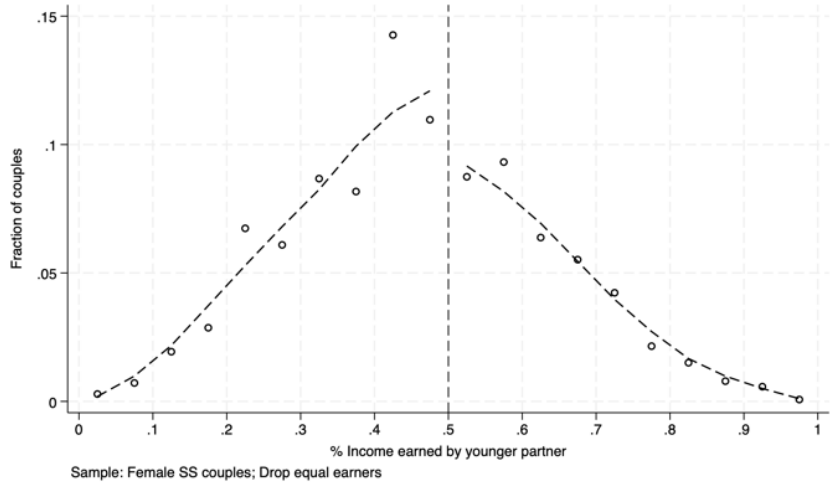

(a) Female same-sex couples

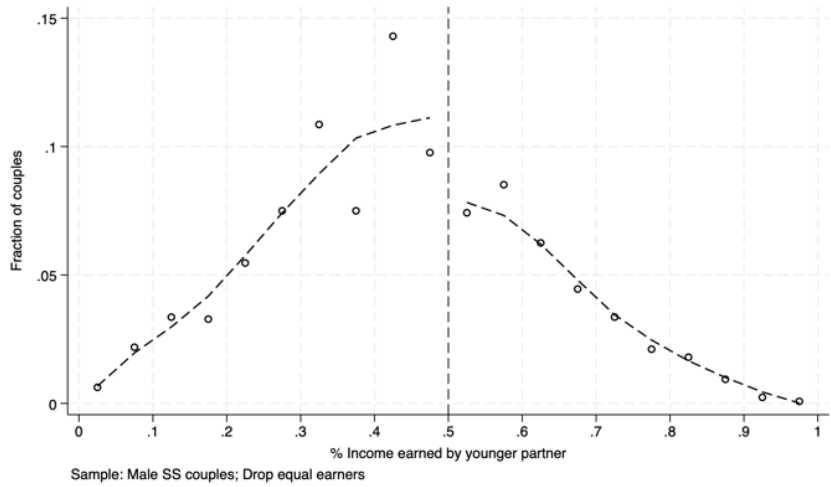

(b) Male same-sex couples

*Notes:* This figure uses data from the 2010 Brazilian Census (10% sample). It presents the distribution of the share of total household labor earnings earned by the younger partner in same-sex couples. Equal earners are excluded. Panel (a) shows female same-sex couples, and Panel (b) shows male same-sex couples.

**Figure C8: Distribution of the relative income earned by the wife under random coupling in Brazil**

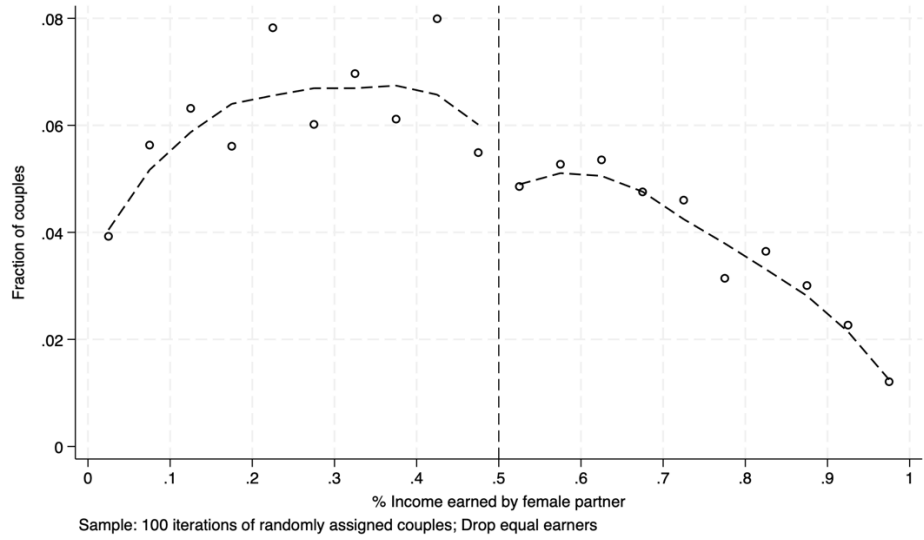

*Notes:* This figure uses data from the 2010 Brazilian Census (10% sample). It presents the distribution of the share of total household labor earnings earned by the wife under simulated random coupling. The figure is based on 100 iterations of randomly assigned different-sex couples. Equal earners are excluded.

**Figure C9: Distribution of households according to relative income in households where the wife is the household head**

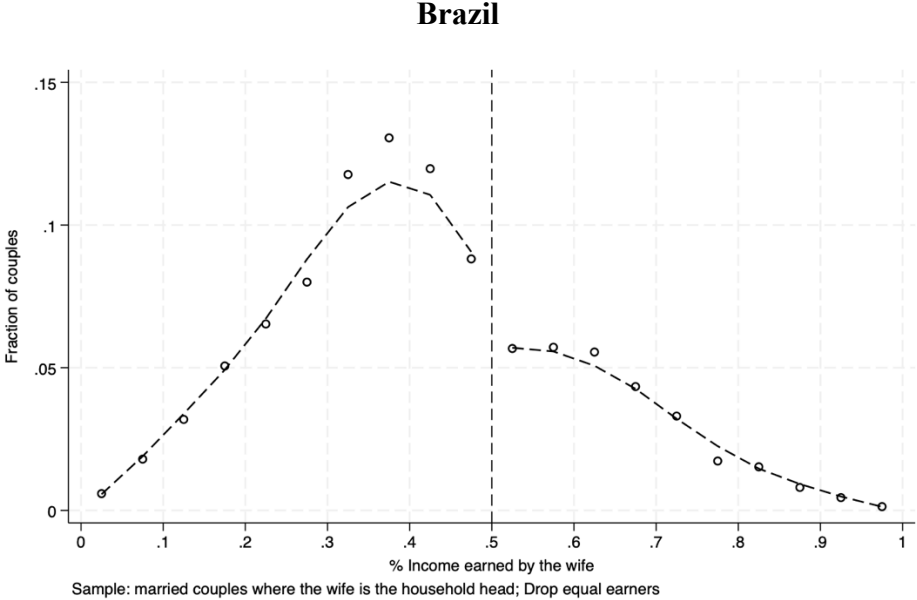

*Notes:* This figure uses data from the 2010 Brazilian Census (10% sample). It shows the distribution of the share of total household labor earnings earned by the wife, restricting the sample to married couples in which the wife is listed as the household head. Equal earners are excluded.

**Figure C10: Distribution of households according to relative income in households where the wife is older than her husband**

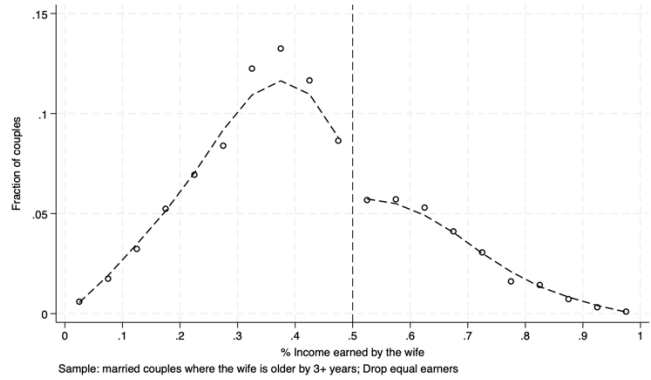

(a) Wife is older by >3 years

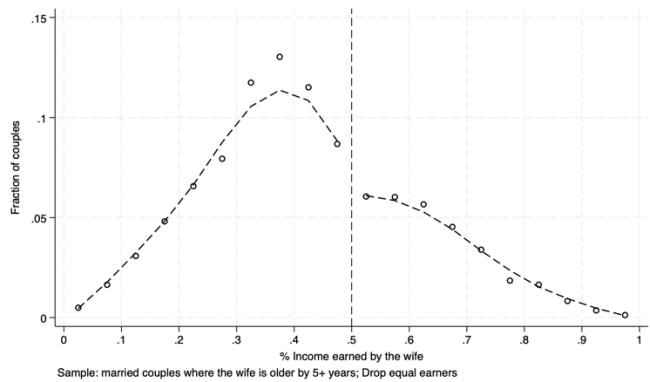

(b) Wife is older by >5 years

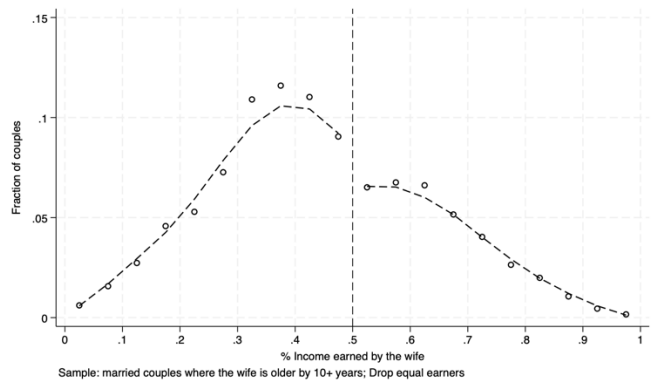

(c) Wife is older by >10 years

*Notes:* This figure uses data from the 2010 Brazilian Census (10% sample). It shows the distribution of the share of total household labor earnings earned by the wife among married couples in which the wife is older than the husband. Panels vary by age difference threshold: more than 3 years (panel a), more than 5 years (panel b), and more than 10 years (panel c). Equal earners are excluded from the sample.

**Figure C11: Heterogeneity by Type of Union in Brazil**

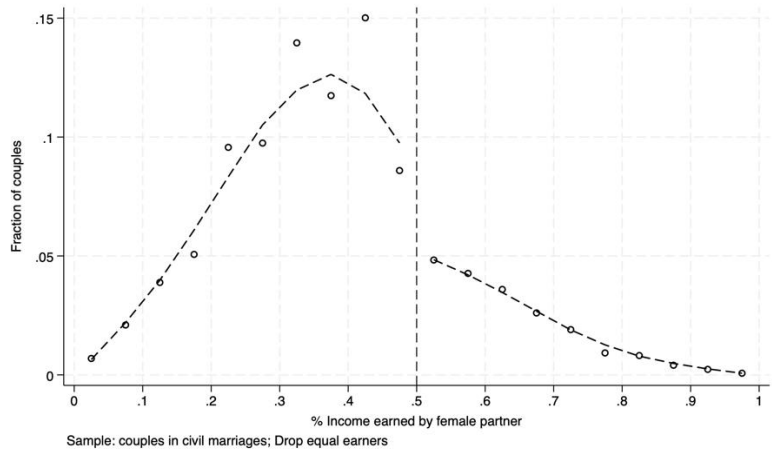

(a) Civil Marriages

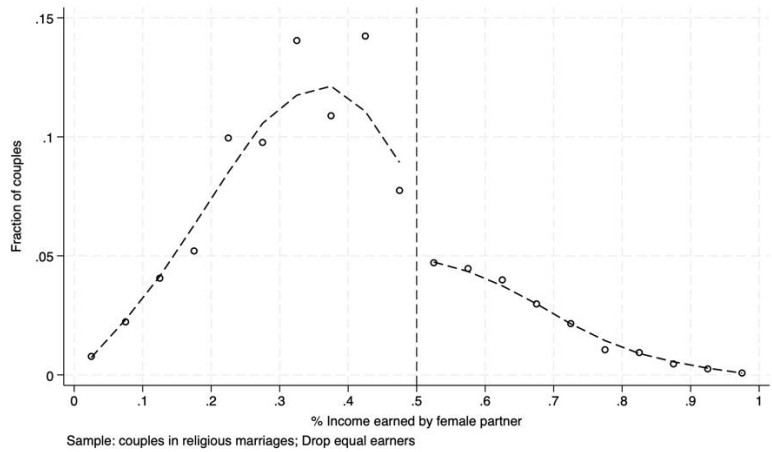

(b) Religious Marriages

*Notes:* This figure uses data from the 2010 Brazilian Census (10% sample). It presents the distribution of the share of total household labor earnings earned by the female partner, separately for married couples reporting civil marriages only (panel a) and those reporting religious marriages, which include both religious only and civil and religious marriages (panel b). Equal earners are excluded from both samples.

**Tables**

**Table C1: McCrary Test for discontinuity in the distribution of the woman's share of total household labor income - Different-Sex Married Partners**

|  | log distance at 0.50001 |
|---|---|
|  | Brazil |
|  | (1) |
| *Sample:* |  |
| All employed couples | -0.250*** |
|  | (0.007) |
| N | 850,593 |
|  |  |
| Drop employers/self-employed | -0.257*** |
|  | (0.008) |
| N | 729,870 |
|  |  |
| Drop same position and occupation | -0.217*** |
|  | (0.008) |
| N | 736,755 |
|  |  |
| Drop employers and self-employed/same position and occupation | -0.220*** |
|  | (0.008) |
| N | 675,389 |

*Notes:* *** $p<0.01$ ** $p<0.05$ * $p<0.10$. Each cell reports the estimated discontinuity in the density of the woman's share of total household labor income at the 50% threshold, using the McCrary (2008) test. The dependent variable is the log difference in the density just above versus just below the 50% threshold. All samples include different-sex married couples in the 2010 Brazilian Census 10% sample. Row 1 includes all employed couples. Row 2 excludes couples in which either partner is an employer or self-employed. Row 3 excludes couples in which both partners work in the same occupation and position. Row 4 applies both exclusions. Standard errors are in parentheses.

**Table C2: McCrary Test for discontinuity in the distribution of the woman's share of total household labor income in 2000 - Different-Sex Partners**

| | log distance at 0.50001 | |
|---|---|---|
| | Brazil Married Partners | Brazil Unmarried, cohabiting partners |
| | (1) | (1) |
| *Sample:* | | |
| All employed couples | -0.096*** | -0.368*** |
| | (0.006) | (0.009) |
| N | 945,179 | 444,796 |
| | | |
| Drop employers/self-employed | -0.106*** | -0.378*** |
| | (0.007) | (0.010) |
| N | 719,965 | 397,710 |
| | | |
| Drop same position and occupation | -0.067*** | -0.337*** |
| | (0.007) | (0.010) |
| N | 883,250 | 395,413 |
| | | |
| Drop employers and self-employed/same position and occupation | -0.072*** | -0.343*** |
| | (0.007) | (0.010) |
| N | 689,550 | 368,308 |

*Notes:* *** $p<0.01$ ** $p<0.05$ * $p<0.10$. Each cell reports the estimated discontinuity in the density of the woman's share of total household labor income at the 50% threshold, using the McCrary (2008) test. The dependent variable is the log difference in the density just above versus just below the 50% threshold. All samples include different-sex couples from the 2000 Brazilian Census. Column (1) restricts the sample to married partners; Column (2) restricts the sample to cohabiting, unmarried partners. Row 1 includes all employed couples. Row 2 excludes couples in which either partner is an employer or self-employed. Row 3 excludes couples in which both partners work in the same occupation and position. Row 4 applies both exclusions. Standard errors are in parentheses.

**Table C3: McCrary test for discontinuities in the distribution of the non-household head/younger partner's share of total labor earnings - Same-Sex Partners, Brazil**

|  | log distance at 0.50001 | |
|---|---|---|
|  | Female S.S. households | Male S.S. households |
|  | (1) | (2) |
| **Panel A: Non-household head partner** | | |
| All dual-earning couples | -0.059 | -0.217 |
|  | (0.151) | (0.179) |
| N | 1,395 | 1,280 |
|  | | |
| **Panel B: Younger Partner** | | |
| All dual-earning couples | -0.135 | -0.072 |
|  | (0.152) | (0.166) |
| N | 1,395 | 1,280 |

*Notes:* *** $p<0.01$ ** $p<0.05$ * $p<0.10$. Each cell reports the estimated discontinuity in the density of the designated partner's share of total labor earnings at the 50% threshold, using the McCrary (2008) test. The dependent variable is the log difference in the density just above versus just below the 50% threshold. The sample includes same-sex dual-earning couples from the Brazilian Census. Panel A defines the designated partner as the non-household head; Panel B defines the designated partner as the younger partner. Column (1) reports estimates for female same-sex households; Column (2) for male same-sex households. Standard errors are in parentheses.

**Table C4: Potential Income and Female Labor Force Participation**

|  | Brazil |
| --- | --- |
|  | (1) |
| *PrWifeEarnsMore* | 0.025*** |
|  | (0.005) |
|  |  |
| Mean of *wifeLFP* | 0.641 |
| *N* | 2,881,649 |

*Notes:* *** $p<0.01$ ** $p<0.05$ * $p<0.10$. Each cell reports the coefficient from a regression of female labor force participation on the probability that the wife would earn more than her husband if both worked full-time. The sample includes married different-sex couples in Brazil. Standard errors are in parentheses.

**Appendix D. Results for Panama**

We replicate our main analyses using the full-count 2023 Panamanian Census. While the data offer national coverage, they include far less detail on household structure, labor market behavior, and marital characteristics than the Mexican and Brazilian censuses. As a result, our analysis is limited to different-sex couples, and we are unable to explore heterogeneity by union type or examine same-sex households.

Despite these limitations– and despite a substantially smaller analytical sample relative to Mexico and Brazil– all main patterns replicate. We continue to find a clear discontinuity in the distribution of the woman's share of household income at the 50% threshold, with magnitudes slightly smaller than those in the Mexican data. The only exception is the division by relative age: unlike in Mexico and Brazil, we do not find stronger discontinuities when the wife is younger than the husband. This exception may reflect noise due to the smaller sample.

This section is organized as follows. Figure D1 show the Gaussian Kernel Density of relative income (Figure A2 for Mexico). Figures D2 and D3 show the relative income distribution in Panama for 2023, 2010 and 2000, respectively (Figures 1 and A4 for Mexico). Figure D4 shows the heterogeneous distributions according to parenting status (Figure 2, Panels (a) and (b)). Figure D5 shows the counterfactual distribution of relative income in Panama under random sorting (Figure A5). Figure D6 shows the distribution for households where the wife is the household head (Figure 3). Figure D7 shows the distribution for households where the wife is older than her husband (Figures 4 and A7).

**Figure D1: Kernel Density of Relative Income – Panama, 2023**

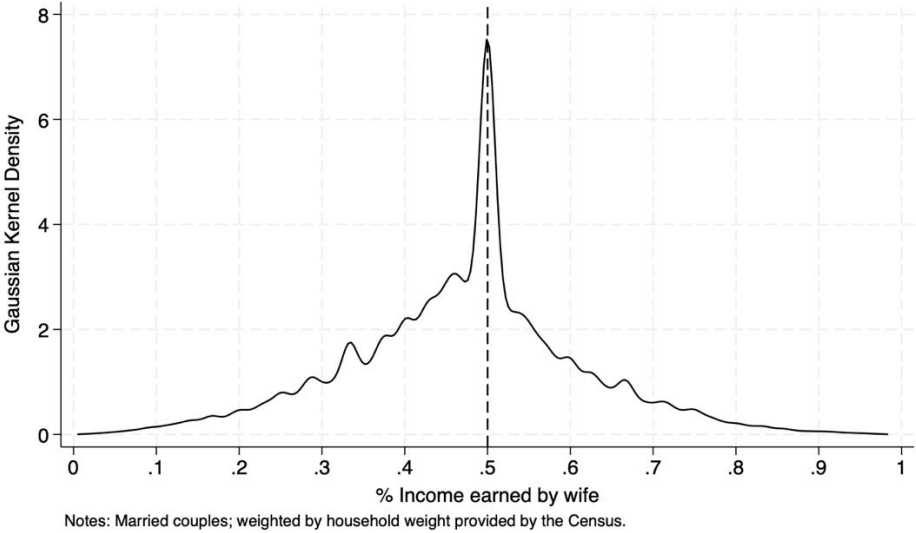

(a) Full sample

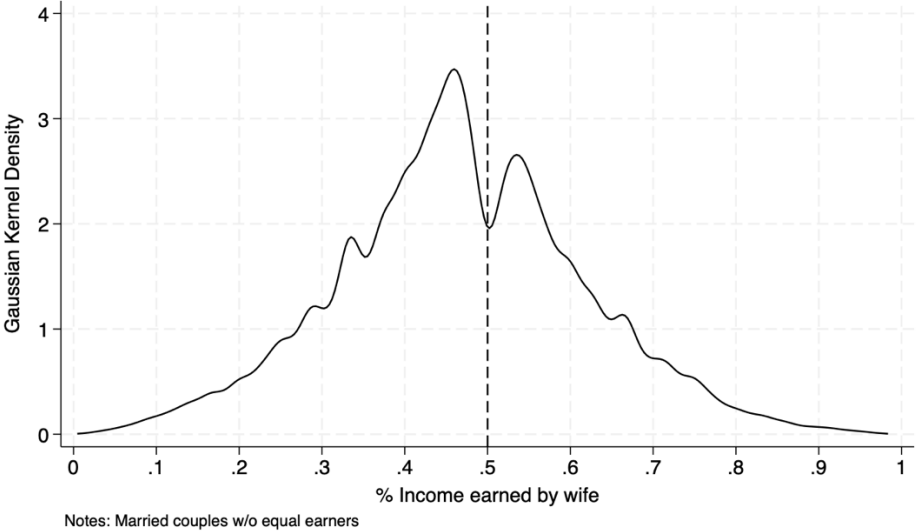

(b) Without equal earners

*Notes:* Sample includes married different-sex couples in the full count 2023 Panamanian census. Panel (a) shows the distribution of the wife's share of household labor income, using Gaussian kernel density and household sampling weights. Panel (b) drops couples with equal earnings.

**Figure D2. Distribution of the share of total household labor earnings by the wife in Panama**

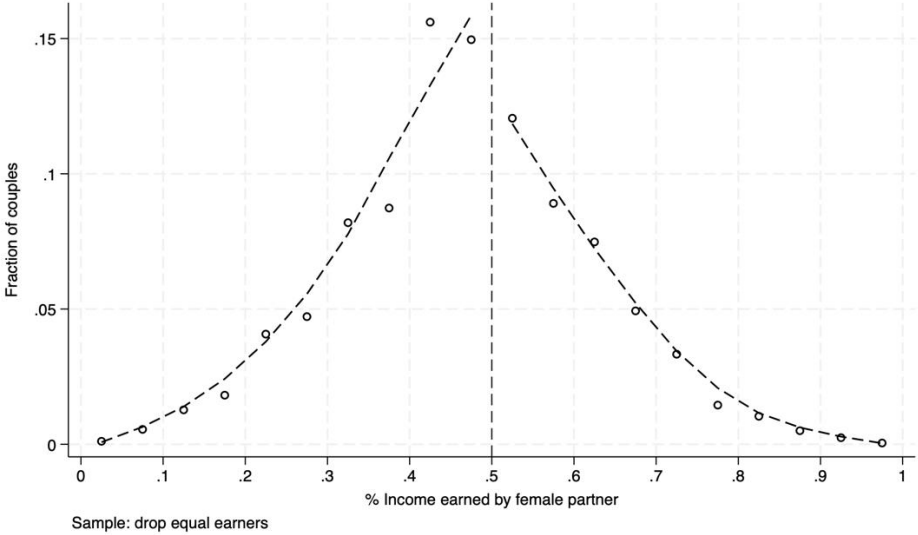

*Notes:* This figure uses data from the 2023 Panamanian Census (full count). It displays the distribution of the share of total household labor earnings earned by the female partner among different-sex couples. Equal earners are excluded from the sample.

**Figure D3: Distribution of the share of total household labor earnings by the wife in Panama in 2000 and 2010**

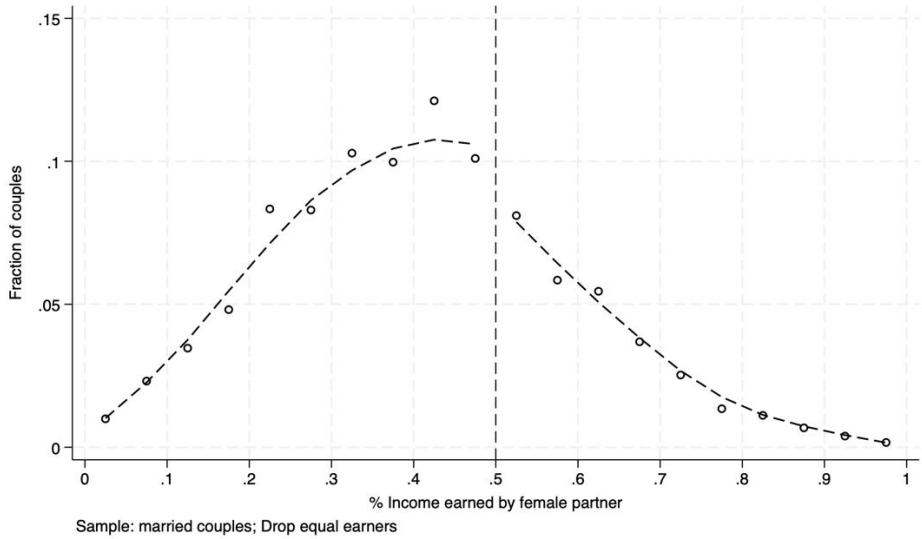

(a) Panama (2000)

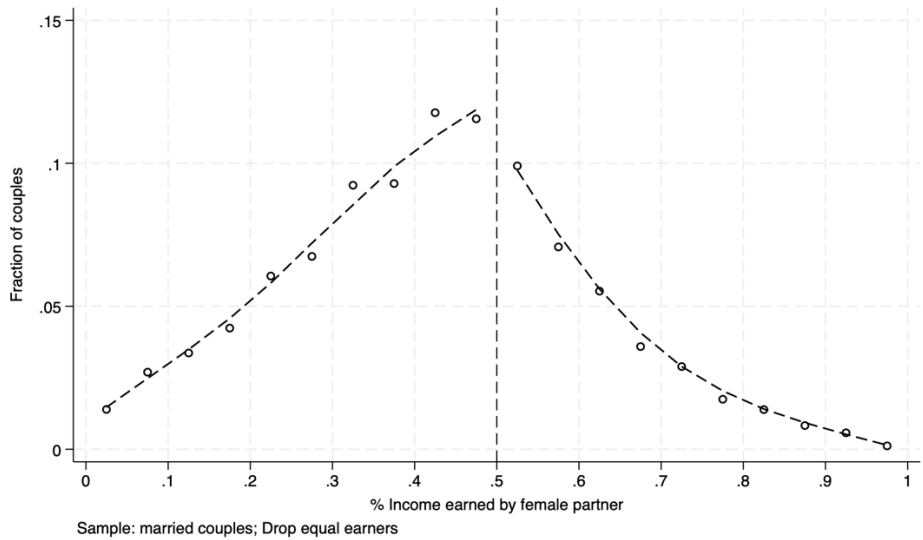

(b) Panama (2010)

*Notes:* This figure plots the distribution of the wife's share of total household labor income using data from the 2000 and 2010 Panamanian Censuses (10% samples). The sample includes married different-sex couples and excludes equal earners. The vertical line marks the 50% threshold in relative income.

**Figure D4: Distribution of couples of relative earnings of female partner according to parenting status in Panama**

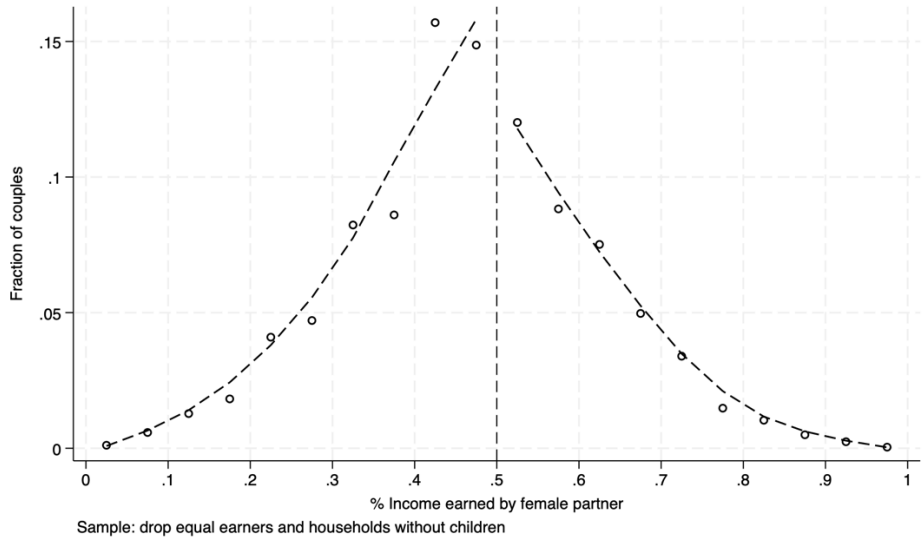

(a) With children

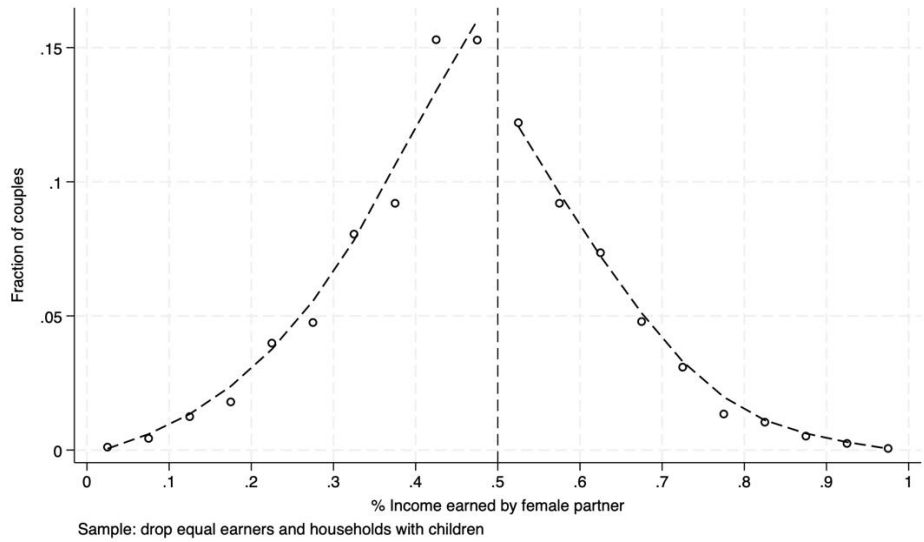

(b) Without children

*Notes:* This figure uses data from the full count 2023 Panamanian Census to show the distribution of the female partner's share of household labor income by parenting status. Panel (a) restricts the sample to households with children, while panel (b) includes only households without children and excludes employers and self-employed individuals. Equal earners are dropped in both panels. The vertical line marks the 50% income share threshold.

**Figure D5: Distribution of the relative income earned by the wife under random coupling in Panama**

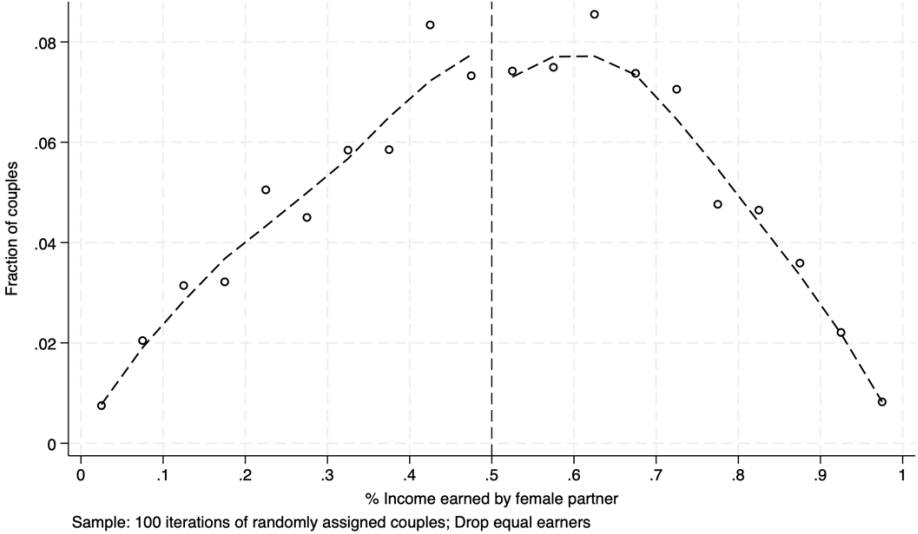

*Notes:* This figure uses data from the full count 2023 Panamanian census. It presents the distribution of the share of total household labor earnings earned by the wife under simulated random coupling. The figure is based on 100 iterations of randomly assigned different-sex couples. Equal earners are excluded.

**Figure D6: Distribution of households according to relative income in households where the wife is the household head**

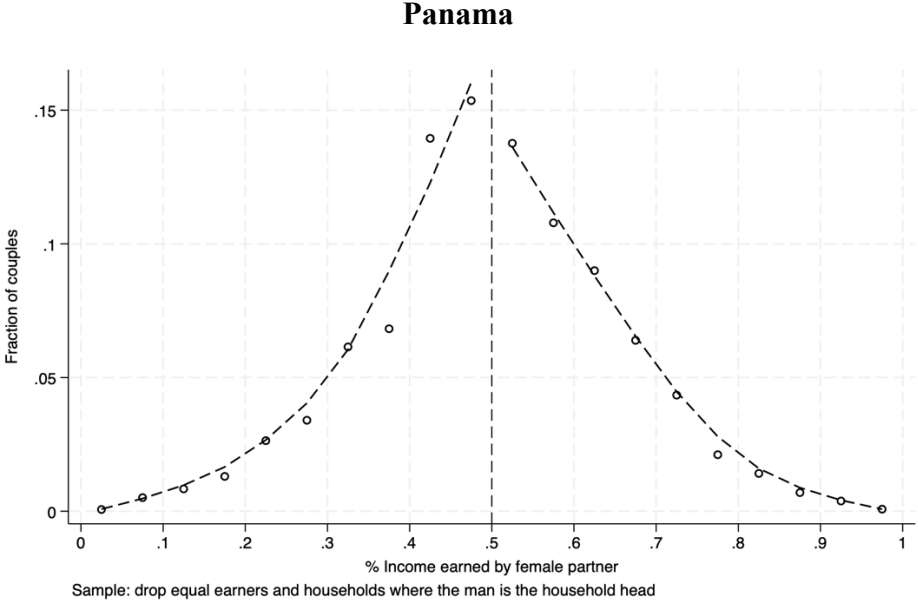

Panama

*Notes:* This figure uses data from the full count 2023 Panamanian Census. It shows the distribution of the share of total household labor earnings earned by the wife, restricting the sample to married couples in which the wife is listed as the household head. Equal earners are excluded.

**Figure D7: Distribution of households according to relative income in households where the wife is older than her husband**

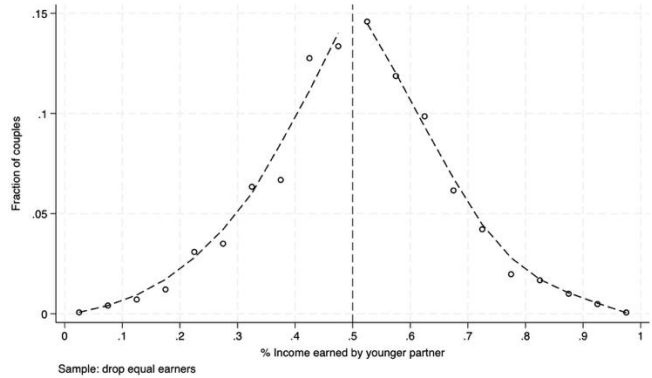

(a) Wife is older by >3 years

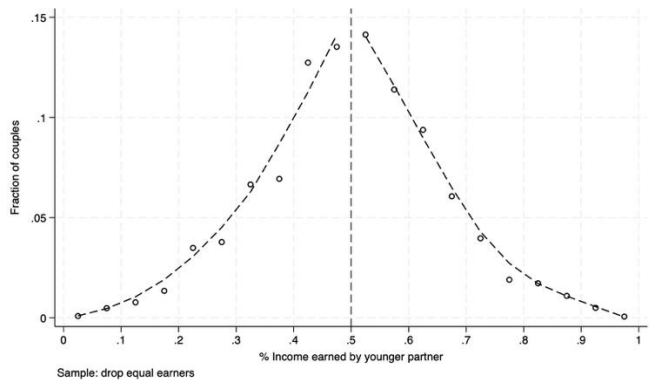

(b) Wife is older by >5 years

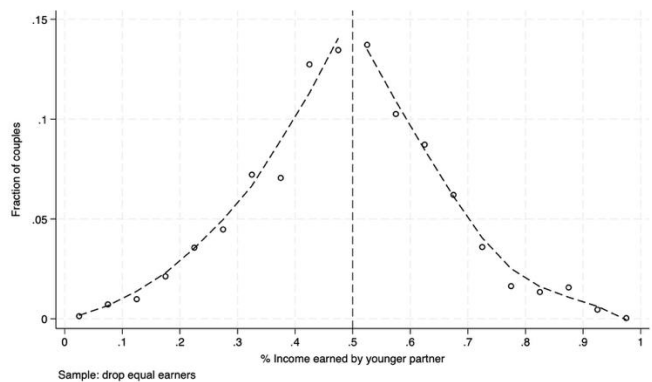

(c) Wife is older by >10 years

*Notes:* This figure uses data from the 2023 full count Panamanian census. It shows the distribution of the share of total household labor earnings earned by the wife among married couples in which the wife is older than the husband. Panels vary by age difference threshold: more than 3 years (panel a), more than 5 years (panel b), and more than 10 years (panel c). Equal earners are excluded from the sample.

**Table D1. McCrary test for discontinuity in the distribution of the woman's share of total labor household earnings – Different-sex married partners – Panama**

|  | log distance at 0.50001 | |
|---|---|---|
|  | Panama Full sample | Panama Excluding equal earners |
|  | (1) | (1) |
| *Sample:* | | |
| All employed couples | -1.306*** | -0.183*** |
|  | (0.011) | (0.013) |
| N | 130,917 | 112,175 |
| | | |
| Drop employers/self-employed | -1.304*** | -0.183*** |
|  | (0.011) | (0.013) |
| N | 130,668 | 111,965 |
| | | |
| Drop same position and occupation | -1.169*** | -0.162*** |
|  | (0.014) | (0.016) |
| N | 85,471 | 75,133 |
| | | |
| Drop employers and self-employed/same position and occupation | -1.166*** | -0.162*** |
|  | (0.014) | (0.016) |
| N | 85,240 | 74,933 |

*Notes:* *** $p<0.01$ ** $p<0.05$ * $p<0.10$. Each cell reports the estimated discontinuity in the density of the woman's share of total household labor income using the McCrary (2008) test. The dependent variable is the log difference in the density just above versus just below 0.50001. Column (1) includes the full sample of different-sex married couples in Panama. Column (2) excludes couples with exactly equal earnings. Each row drops additional subsets: self-employed and employer couples, couples in the same position and industry, or both. Standard errors are in parentheses.